\documentclass[pra,
aps,
10pt,
twocolumn,
floatfix,
superscriptaddress,
notitlepage,
longbibliography]{revtex4-1}
\usepackage[T1]{fontenc}
\usepackage[normalem]{ulem}
\usepackage{graphicx, color, graphpap}% Include figure files
\usepackage{enumitem}
\usepackage{amssymb}
\usepackage{multirow}
\usepackage[colorlinks=true,citecolor=blue,linkcolor=magenta]{hyperref}
\usepackage{verbatim}
\usepackage{mathtools}
\usepackage[toc,page]{appendix}
\usepackage{bm}
\usepackage[linesnumbered,ruled,vlined]{algorithm2e}
\usepackage[noend]{algpseudocode}
\SetKwInput{kwInit}{Init}
\long\def\ca#1\cb{} %Use for commenting out: \ca...\cb

\usepackage{shorts}
\begin{document}
\title{Automatic re-calibration of quantum devices by reinforcement learning}  \author{T. Crosta}
\affiliation{Computer Vision Center (CVC), 08193 Bellaterra (Cerdanyola del Vallès), Spain}
%\email{tomycrosta@gmail.com}
\author{L. Rebón}
%\email{lrebon@gmail.com}
\affiliation{Instituto de F\'isica La Plata (IFLP), CONICET - UNLP, and Departamento de Ciencias Básicas, Facultad de Ingeniería, Universidad Nacional de La Plata (UNLP), La Plata 1900, Argentina}
%\affilitation{Departamento de Ciencias Básicas, Facultad de Ingeniería, Universidad Nacional de La Plata, La Plata 1900, Argentina}
\author{F. Vilariño}
%\email{fernando@uab.cat}
\affiliation{Computer Vision Center (CVC), 08193 Bellaterra (Cerdanyola del Vallès), Spain}
\affiliation{Department of Computer Science, \\  Universitat Autònoma de Barcelona (UAB), 8193 Bellaterra (Cerdanyola del Vallès), Spain.}
\author{J. M. Matera}
%\email{matera@fisica.unlp.edu.ar}
\affiliation{IFLP-CONICET, Departamento de F{\'i}sica, Facultad de Ciencias Exactas, Universidad Nacional de La Plata, C.C. 67, La Plata 1900, Argentina}
\author{M. Bilkis}
\email{mbilkis@cvc.uab.cat}
\affiliation{Computer Vision Center (CVC), 08193 Bellaterra (Cerdanyola del Vallès), Spain}
\date{\today}

\begin{abstract}
During their operation,  due to shifts in environmental conditions, devices undergo various forms of detuning from their optimal settings. Typically, this is addressed through control loops, which monitor variables and the device performance, to maintain settings at their optimal values. Quantum devices are particularly challenging since their functionality relies on precisely tuning their parameters. At the same time, the detailed modeling of the environmental behavior is often computationally unaffordable, while a direct measure of the parameters defining the system state is costly and introduces extra noise in the mechanism.
In this study, we investigate the application of reinforcement learning techniques to develop a model-free control loop for continuous recalibration of quantum device parameters. Furthermore, we explore the advantages of incorporating minimal environmental noise models. As an example, the application to numerical simulations of a Kennedy receiver-based long-distance quantum communication protocol is presented.
\end{abstract}

\maketitle

\section{Introduction}
Calibrating an experimental apparatus is a primitive and ubiquitous task in most areas of science and technology. In turn, sensor and detector devices constitute the way to extract information about the environment surrounding us and better understand reality via further post-processing of the acquired data. Thus, fully calibrating experimental devices is a primordial task and, in turn, an active research topic~\cite{cimini2021calibration, cimini2024variational, ren2020sensor,vernuccio2022artificial, ono2016calibration, zhao2019online, haitjema2020the, cimini2019calibration, zhang2021machine, bilkis2020real, bilkis2021reinforcement}. 
\begin{comment}
    li2024enhancing, gao2015a, asif2022realtime, yang2024temperatureautomated, vajs2023datadriven, cao2020realtime, chen2022a, papafotis2021magnetic, harms2016consistent, masinde2014an, lin2023an, valensise2021deep, jung2017multiobjective, liu2023machinelearningbased,martins2023online,
\end{comment}
In this manuscript, we study the recurrent calibration of devices whose deployment environment is challenging to be modelled. Examples of this are scenarios that heavily vary with time in a way that is hard to predict, \textit{e.g.} turbulent atmosphere~\cite{Dequal2020, Andrews2005, Pirandola2021a, Pirandola2021,Vasylyev2011,Vasylyev2017}, hydrological models~\cite{jung2017multiobjective,KAVETSKI2006173} or non-isolated magnetometers~\cite{papafotis2021magnetic,cao2020realtime} among many others. For such settings, where state-of-the art technology is being used to push forward the boundaries of scientific discoveries at a considerable resource overhead, it is of utmost importance to develop techniques that are ready to adapt the device configuration to the experimental condition at hand. In this regard, a plethora of artificial-intelligence techniques have recently been developed in the context of sensor calibration~\cite{
fallani2022learning, fiderer2021neural, Fiderer2018qchaos,cimini2019calibration, lee2021quantum, nolan2021a,ban2021neuralnetworkbased, chen2022a, rambhatla2020adaptive, cimini2021calibration, cimini2024variational}, change-point detection~\cite{Sentis2016cgp, Fanizza2023ultimate, Sentis2017exact} and malfunctioning device identification~\cite{Wozniak2023quantumanomaly, baker2022quantumvariational, Guo2023quantumalgo, Llorens2023quantummuti,Michalis20218Identification, Liu2018QML}. 

Our main contribution is to provide a framework for re-calibrating quantum devices. Based on this, we present an automatic re-calibration method, and showcase it in a quantum-measurement scheme crucial for long-distance communication by laser pulses, \textit{e.g.} satellite-ground or optical-fiber communication. Overall, the success of most machine-learning (re)calibration schemes considered in literature rely either on perfect knowledge of devices' functioning conditions, or in access to huge amount of data for training purposes. Such assumptions constitute a double-edged sword when deploying the device on (potentially adversary) experimental conditions: while correct configurations can be granted if the machine-learning model was trained on data resembling the experimental conditions at hand, with high probability the device will otherwise remain off-calibrated.

Here, we depart from such a notion of similarity between training and deployment scenarios, by considering a hybrid scheme consisting on a pre-training round complemented with a reinforcement-learning stage. The latter fine-tunes the configuration, so the device can be adapted to the specific (and potentially unexplored) experimental conditions at hand; this is done by modifying device controls, as shown in Fig.~\ref{fig:fig1}.

The success of our method hinges on the capabilities of devising an approximate model of setting's dependence with respect to changes in its surroundings (which we indistinctly call environment). Such approximate model is to be thought as a simplified description of the environment, \textit{e.g.} captured by very few variables. While not expected to be fully accurate --- does not retrieve the exact device configuration for each specific experimental condition---, it shall be thought as an ansatz for controls initialization. The accuracy of this initialization relies on the capacity of the approximated model to capture relevant features of device behavior given the experimental condition at hand. As a rule of thumb, the more complex such ansatz is made, the more accurate the description is expected to be, although a trade-off rises. While more complex models tend to be more data-consuming until reaching optimal calibrations, tailoring the model to specific experimental conditions will inherently induce a bias towards a sub-set of deployment scenarios. Thus, the goal of the pre-training round is to suggest control initialization values by using a small number of quantities that can easily be estimated out of a few experiments. The control values are then improved by means of a complementary reinforcement-learning method, which adapts the control values to the specific experimental condition in a model-free way. On top of the calibration mechanism, the value of a de-calibration witness is continuously monitored during deployment, which allows the agent to experimentally detect that the device entered an off-calibration stage, and thus re-initiate the calibration process. 

This work is a further step towards developing a fully automatic re-calibration of quantum detectors through machine-learning techniques. Importantly, we remark that neither the framework nor the method is specially biased towards the quantum realm, and can potentially be applied to other control problems beyond the quantum-technology scope.

The manuscript is structured as follows. In Sec.~\ref{sec:framework} we present our re-calibration framework and described our method. In Sec.~\ref{sec:results} we numerically analyze the performance of our re-calibration method in a paradigmatic long-distance quantum communication setting. Conclusions and future work are outlined in Sec.~\ref{sec:chau}
 
\section{The re-calibration framework}\label{sec:framework}
\begin{figure}[t!]
\centering
    \includegraphics[width=0.45\textwidth]{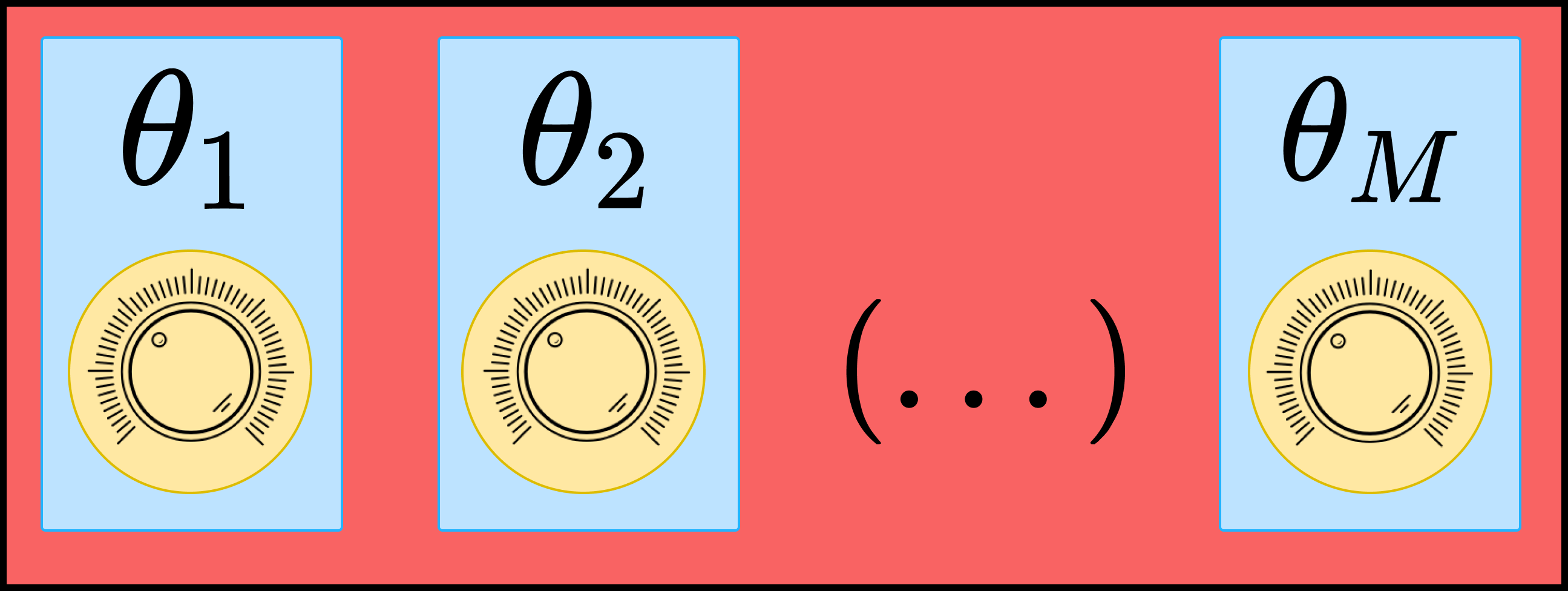}
    \caption{We depict a device that needs to be calibrated. Here, the apparatus is controlled by different knobs defined by values $\bm{\theta} = \llaves{\theta_1, ..., \theta_M}$, and the aim is to tune such parameters in a way that the device is configured to optimally operate under experimental conditions $\cE$.}
    \label{fig:fig1}
\end{figure}

We consider a device whose controls are defined by continuous parameters $\bm{\theta} = \llaves{\theta_1,...,\theta_M}$. As shown in Fig.~\ref{fig:fig1}, our setting is a black-box device controlled by different knobs $k=1,...,M$ each associated to a control value $\theta_k$. In the following, we define several quantities of interest.

\textit{Device configuration}. A fixed set of parameter values $\bm{\theta}$ completely defines a device configuration.

\textit{Score function}. The quality of a device configuration $\bm{\theta}$ is evaluated by a score function $S_{\cE}(\bm{\theta})$. The value of the score function can be estimated---during a calibration stage--- by means of $N$ repeated experiments; each experiment $i$ involves a quantum measurement and leads to a measurement outcome $\bm{n}_i$, whose value is generally of stochastic nature. Here, the full underlying model needed to describe outcome probability distributions is denoted by $\cE$, and generally involves an accurate description of noisy channels present in the setting at hand.

\textit{Effective score function}. The underlying model $\cE$ is generally inaccessible to the calibrating agent and hence shall assume to be unknown to her. This is motivated by the fact that: \textit{(i)} time-varying deployment conditions can fundamentally be hard to model, and \textit{(ii)} even in the case of having full control of experimental conditions, quantum channel-tomography comes with a considerable sample overhead, implying that the number of experiments and parameters required to reach near-optimal environment modelling (plus device calibration) would grow exponentially or be otherwise constrained to specific scenarios ~\cite{nielsen00,Branderhorst2009simplified, Shabani2011efficient, dimario2021channelnoise}. On the contrary, we do assume that a certain relationship exists between the \textit{true} score function $\sct$ and its effective version $\sca$, \textit{e.g.} the effective model $\ctE$ used by the agent is indeed able to capture certain relevant features of the score function. Effective models should be thought as an enhanced \textit{control initialization} strategy. This notion applies for the case in which the device enters an off-calibration stage, and new control values should be found. 
\begin{figure}[t!]
    \centering
    \includegraphics[scale=0.37]{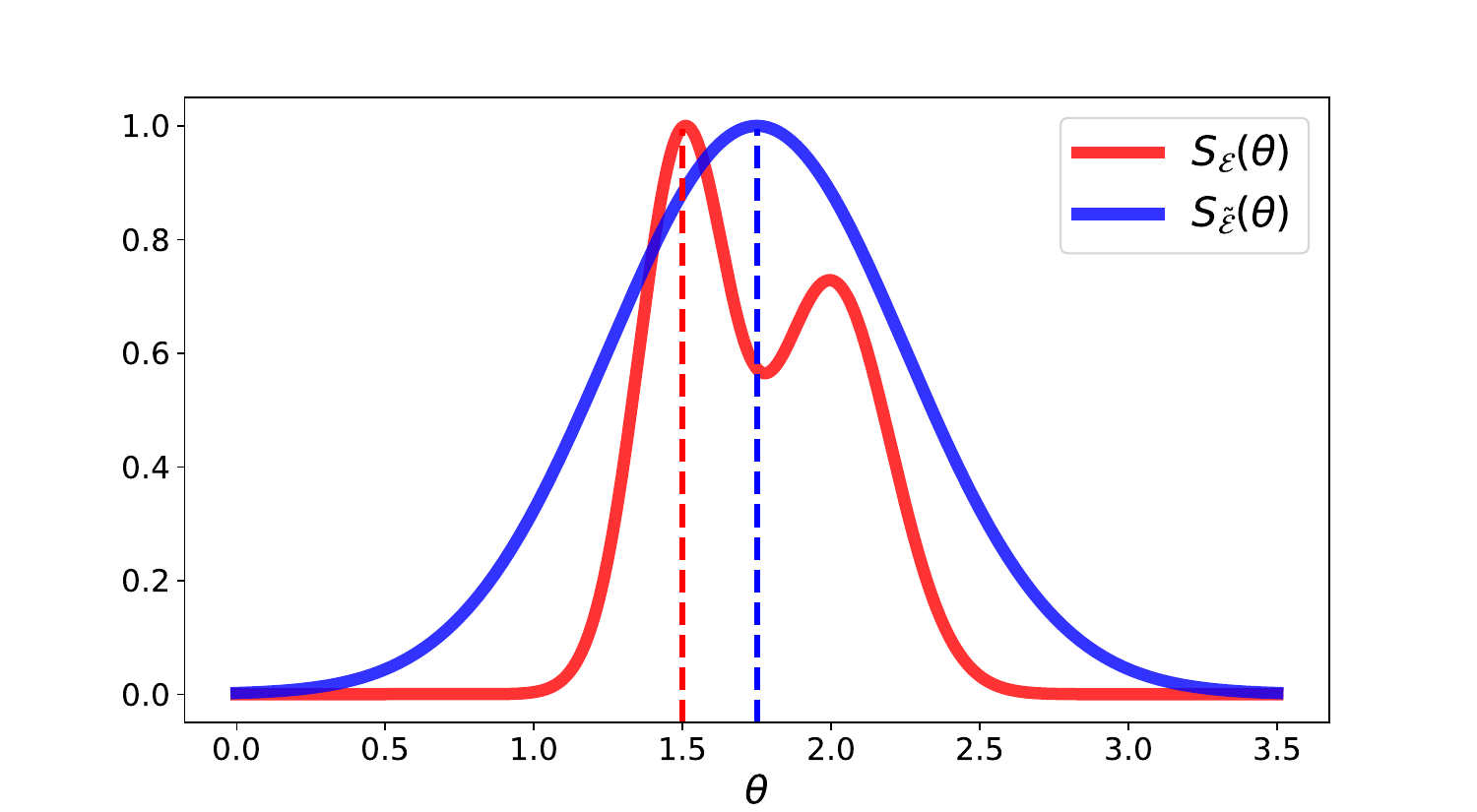}
     \includegraphics[scale=0.37]{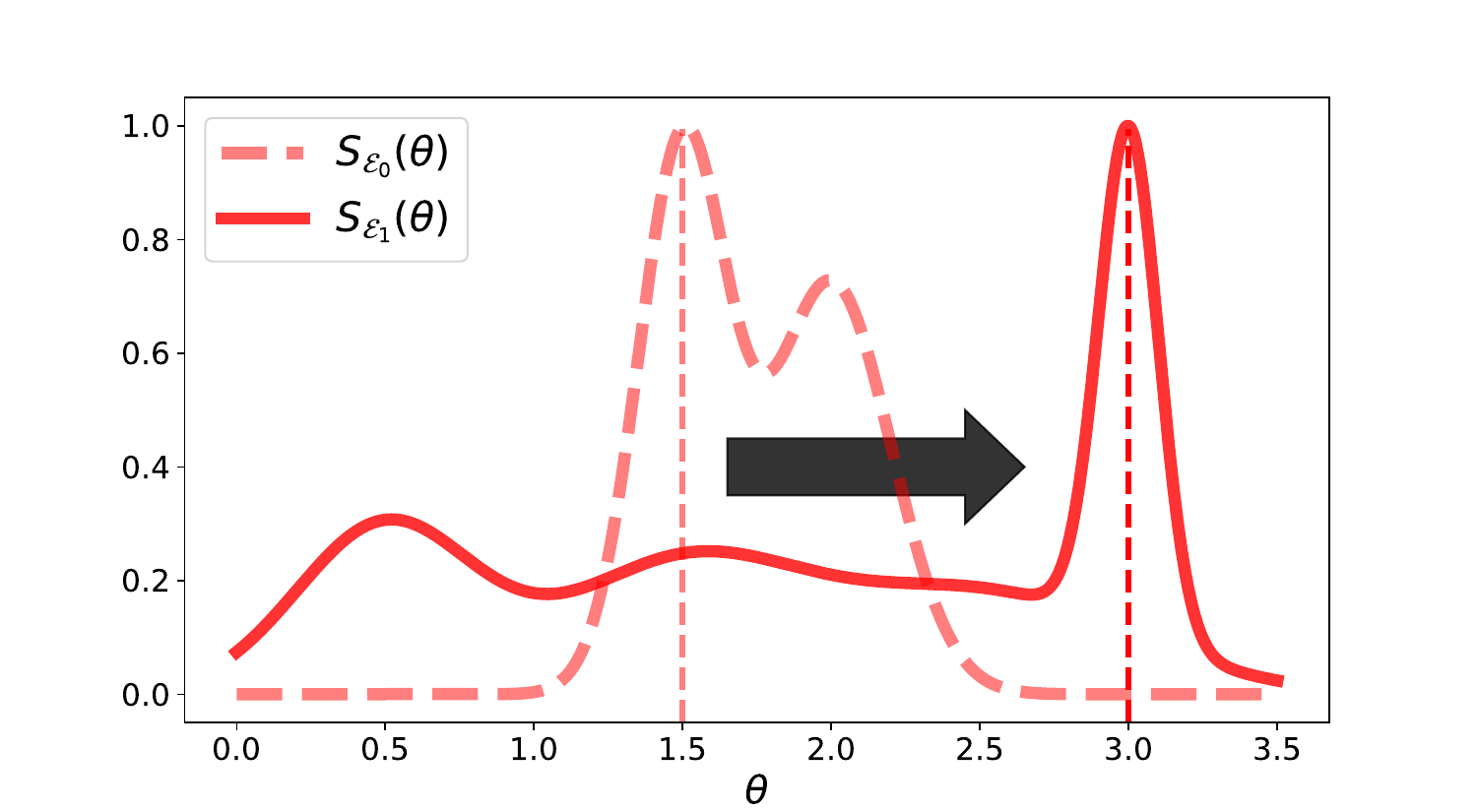}
    \caption{Single-parameter device example. \textit{Top panel}: the optimal calibration score $\sct$ is shown (dashed-red vertical line), and its \textit{effective} value $\sca$ (blue-dashed vertical line); while sub-optimal, this value is further fine-tuned by means of a model-free scheme (see main body). \textit{Bottom panel}. We show score functions $S_{\mathcal{E}_0}$ and $S_{\mathcal{E}_1}$ before and after a change-point occurs in the environment. As a consequence, the device optimally configured under $\mathcal{E}_0$ needs now to be re-calibrated to the new optimal configuration for $\mathcal{E}_1$. }
    \label{fig:fig2}
\end{figure}

\textit{Reinforcement Learning (RL)}. The setting described above can be framed in the RL language~\cite{Sutton2018, Dawid2022modern, bilkis2020real,Borah2021measurement, fallani2022learning,Briegel2012projective}, where an agent repeatedly interacts with an environment in order to maximize a reward function, during different episodes.Here, at $i^{\text{th}}$ episode (experiment), the agent selects parameter values $\bm{\theta}$, observes measurement outcomes $\bm{n}_i$, and finally post-processes them in order to provide a \textit{claim} for the underlying task the quantum device is used for. Based on the accuracy of this final action, the agent is given a reward signal, which uses to improve its estimate on how valuable the decisions performed were. In RL, this is captured by the so-called \textit{state-action value-function} $Q_\pi(s,a)$, standing for the expected reward when departing from state $s$ and taking action $a$ (\textit{i.e.} either selecting parameters $\bm{\theta}$ or providing a claim based on the outcomes acquired~\cite{Sutton2018,bilkis2020real}), and following decision criteria --- or policy--- $\pi$. For an optimal decision criteria (\textit{e.g.} optimal device usage) the agent shall choose configuration $\bm{\theta}^*$ leading to a maximum score $\sct$. Nonetheless, since $\cE$ is not available to the agent, value functions need to be estimated out of several experiment repetitions. Importantly, agent's strategy is optimized solely based on the rewards acquired during learning. Here, not only such rewards are a way to estimate value-functions, but also serve as a lighthouse for the agent to navigate the decision landscape, allowing a \textit{model-free} calibration of the device. We provide further details on how such model-free calibration work in Appendix~\ref{chap:append}.

As an example, we consider a single-control device, whose score function $\sct$ is schematized in Fig.~\ref{fig:fig2}. A model-free agent would initially set the parameter $\theta$ at random and consequently estimate its score function out of repeated experiments. On the contrary, keeping an effective model $\ctE$ can readily help the agent to improve such initialization strategy. Here, the agent's internal model $\sca$ serves as an ansatz for the underlying behavior of score $\sct$ w.r.t. the control $\theta$. Intuitively, the internal model $\ctE$ is expected to be easier to estimate out of few experiment repetitions.

\begin{comment}
The above definitions capture the performance of a calibrating agent choosing device configuration $\bm{\theta}$ out of several experiment repetitions, via a pre-defined control strategy. Importantly, the agent should balance between exploring new configurations and exploiting valuable and known ones, obtained from empirical score-function estimations. Such a trade-off is captured by the reinforcement-learning framework, which in turn provides a plethora of control search strategies to tackle the (re)calibration problem.
\end{comment}

RL methods have recently been applied to a wide variety of quantum technology scenarios, among them calibrating a quantum communication setting~\cite{bilkis2020real,bilkis2021reinforcement,Wallnofer2020machine, Cui2022quantum, Rengaswamy2021belief,Piveteau2022quantummessage, Cristian2022Sample}, optimizing quantum pulses~\cite{Sivak2022model, Niu2019universal,fallani2022learning, Fosel2018RL}, quantum gated-circuit layout~\cite{Altmann2023challenges, Nagele2023Optimizing}, and even graph-processing applications~\cite{Skolik2022Equivariant}, to name a few. However, little has been studied on the capabilities of the learning model to \textit{adapt} the calibration to changes in the environment $\cE$ happening while the device is being used, \textit{e.g.} in the \textit{deployment stage} (find however some examples falling in this category in  Refs.~\cite{Khandelwal2022enhancing, Zhou17magnetic}). In Fig.~\ref{fig:fig2} (down) we exemplify how a change in the environment would affect the score function, requiring a recalibration. In order to detect the new landscape, we can consider that the new observations will be different from the ones predicted by the previous exploration. Thus, having indications about changes in the environment even during off calibration stages.

\textit{De-calibration witness}. In order to realize that a change occurred in the environment, the agent must rely on an experimentally accessible quantity, which we define as de-calibration witness and denote with $\Wd$. By monitoring the behavior of $\Wd$ across different experiments, the agent can readily detect whether a change-point occurs in the environment and thus re-start the calibration routine if anomalies are detected. Examples of potential de-calibration witnesses are estimates of the outcome probabilities, which the agent can straightforwardly construct from the information gathered across experiments.

\vspace{0.5cm}
\textit{Automatic re-calibration}. The definitions outlined above set a framework to analyze the recurrent calibration of a device. We now turn to describe our automatic re-calibration method, which makes use of effective score functions, RL routines and de-calibration witnesses. Here, we picture a scenario where the device is to be initially calibrated and, while it is being deployed, the device  enters an off-calibration stage which needs to be compensated.

The quality of a given configuration is measured by a score function --- which in turn depends on the current experimental conditions---; the maximum of the latter quantity encodes the solution to the problem for which the device is being used for.

For instance, in a communication setting, the device configuration is defined by the encoding-decoding strategy (\textit{e.g.} the quantum measurement performed to decode information out of the incoming signal), and the score function is given by the success probability of the protocol. Alternatively, in variational quantum computing applications~\cite{Cerezo2021, NISQreviewAlba}, \textit{i.e.} the VQE algorithm~\cite{Peruzzo2014quantum}, the device configuration is defined by the free parameters of the parametrized quantum circuit and the score function is given by the energy landscape, which needs to be estimated out of several repetitions of an experiment. 

The initially-optimal configuration can be attained by model-free RL schemes~\cite{bilkis2020real,Sivak2022model}, \textit{i.e.} trial-and-error learning mechanisms. In this approach, one typically estimates the score function out of the rewards acquired for each device configuration, \textit{i.e.} empirical estimation of value functions (see Appendix~\ref{chap:append}). Here, we depart from this concept by initializing the value function estimates to a surrogate quantity, defined by the effective score function $\sca$; such quantity is to be estimated out of few experiment repetitions, and serves as an ansatz for which score value is assigned to a given device configuration (see Fig.~\ref{fig:fig1} for a schematic representation). The usage of effective score values is motivated by the fact that experimental conditions might not dramatically differ from the ideal case. For example, the VQE energy landscape shall preserve certain similarities between a noiseless scenario and a noisy one, assuming the noise strength is sufficiently low. From this informed initialization of value-function estimates, we then exploit the model-free features of traditional RL algorithms, which allows the agent to fine-tune the device configuration, adapting it to specific deployment conditions.

The mechanism described above constitutes the \textit{calibration stage}, in which the actions performed by the calibrating agents can be rewarded according to their accuracy/correctness. 

With the initial calibration task accomplished, the device is then deployed, \textit{e.g.} used without the necessity of rewarding the agent. As experiments proceed, it is to be expected that the device undergoes a de-calibration, \textit{e.g.} experimental conditions might eventually vary. In order to detect such a change occurs, the agent controls the de-calibration witness $\Wd$ ---for example measurement outcome probabilities ---, which is used by the specific change-point detection protocol the agent keeps. Thus, by monitoring $\Wd$, the agent can detect that the device entered into a de-calibration stage, \textit{e.g.} the deployment conditions have changed. As a consequence, the new optimal configuration is a different one, and a re-calibration is carried out. This is done similarly to the initial calibration stage: the effective model configuration landscape is estimated out of few experiments, and the model-free RL algorithm is then used to adjust the configuration to the new optimal one. 

The effective model used by the agent $\sca$, along with the de-calibration witness $\Wd$ and the RL algorithm (\textit{e.g.} search strategy, value functions, reward definitions), define a re-calibration strategy. However, each of the strategy components requires the agent to pre-set a number of \textit{hyperparameters}. Among them, the number of experiment repetitions needed to estimate effective-model configuration landscape (which we denote as $N_{\text{eff}}$), the number of experiments needed to fine-tune the configuration using a RL method, denoted as $N_{rl}$, the undecision region for which values that take $\Wd$ will not lead the agent to re-activate the calibration routine, and the parameters defining the behavior of the RL routine, whose nature depends on the particular algorithm used. In order to help with notation, we will comprise all such parameters by $\bm{\xi}$; in Algorithm~\ref{alg:(E_Q)-learning} a pseudo-code of our re-calibration method is provided.
\begin{algorithm}[t]\label{alg:(E_Q)-learning}
  \DontPrintSemicolon
  \SetAlgoNoEnd
  \SetKwInOut{Input}{input}\SetKwInOut{Output}{output}
  \Input{$\sca$, $\Wd$, $\bm{\xi}$, $RL$-algorithm}
  \Output{$\bm{\theta}^*$ (optimal configuration)}
  \texttt{Calibration stage} by $\sca$\;
  \texttt{Fine-tuning} by $RL$\;
  \texttt{Deployment stage}\;
  \While{$\Wd$ retrieves \textit{normal}}{
  deploy device \;
  \If{$\Wd$ retrieves \textit{anomaly}}{return to step 1}
}
\caption{Automatic re-calibration method.}
\end{algorithm}

In the following, we showcase the re-calibration method introduced above in a canonical example for long-distance classical-quantum communication. We stress that our method can be applied to a wide variety of scenarios, not necessarily constrained to the quantum technology realm.

\section{Illustrative example and numerical development}\label{sec:results}
As an application example, we consider the binary coherent-state discrimination, which is a primitive used in long-distance classical-quantum communication. The usage of quantum resources is expected to boost long-distance communication rates~\cite{Banaszek2020,Rosati16b} and provide unconditional security~\cite{Pirandola2019}. Optimally performing quantum-state discrimination is of utmost importance to reach capacity rates~\cite{wildebook, Nasser2017polar}, and the binary coherent-state discrimination problem currently stands as a canonical problem both from a theoretical point of view~\cite{helstromBOOK,Assalini2011revisting,Zoratti2021agnostic,Takeoka2005Implementation}, as well as an experimental one~\cite{Cook2007,Sych2016practical,becerra2011mary,Dolinar1973,Dequal2020,Andrews2005,Usenko2012a,Pirandola2021,Vasylyev2011,Vasylyev2017,DiMario2022, bilkis2021reinforcement}. The interest on this problem lies on the fact that the optimal quantum measurement to be done by the receiver can be implemented in a sequential way, combining linear optical operations and feedback operations, which constitutes an experimentally friendly setting.

In the long-distance communications scenario, the sender encodes a bit $k=0,1$ in the phase of a quantum coherent-state $\ket{(-1)^k \alpha}$, which is sent to the receiver, \textit{e.g.} the signal is prepared in an orbital space (satellite) station, travels through the atmosphere and arrives to a receiver, in a ground-earth station. The latter performs a binary-outcome quantum measurement, leading to measurement outcome $n\in\llaves{0,1}$. With this information, the receiver provides a guess $\hat{k}$ on the value of the bit transmitted, and the quality of such protocol is given by the success probability. Such a quantity represents the score function $\sct$ introduced in Sec.~\ref{sec:framework}, and depends on the intensity $|\alpha|^2$ of the transmitted states, the quantum channel acting over which the communication takes place, and the specific quantum measurement that is performed by the receiver.

\begin{figure}[t!]
    \centering
\includegraphics[width=0.45\textwidth]{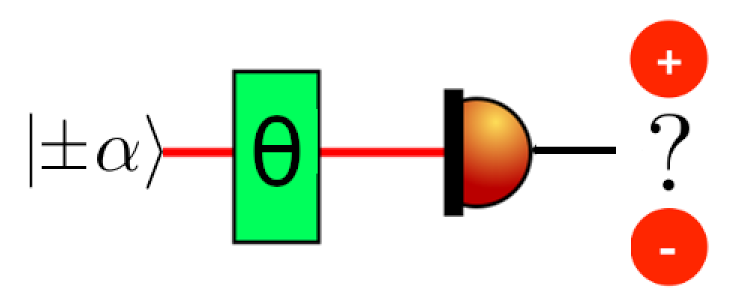}
    \caption{Diagram of a Kennedy receiver; this consists in applying a displacement $\theta$ to the incoming signal and measure it with an on/off photo-detector.}
    \label{fig:kennedy_receiver}
\end{figure}

Among all possible quantum measurements that the receiver can implement, we will here focus on the so-called Kennedy receiver~\cite{Kennedy1973a}, which consists on displacing the incoming signal by a value $\theta$ and measuring the resulting state via an on/off photo-detector, as schematized in Fig.~\ref{fig:kennedy_receiver}. While the optimal quantum measurement is given by the so-called Dolinar receiver~\cite{Dolinar1973,Assalini2011revisting}, which involves complex conditional measurements ultimately leading to difficulties in experimental implementations~\cite{Cook2007, Sych2016practical}, the Kennedy receiver can readily beat the standard quantum limit~\cite{RevGauss} and essentially constitutes the main building block of the former one. 

Thus, in this example the device configuration is defined by \textit{(i)} the parameter $\theta$ in the displacement operation, and \textit{(ii)} a guessing rule which associates the measurement outcome $n$ to the guessed value of the initially transmitted bit $k$. We note that access to the score value (success probability) is granted only in cases where the transmission channel and device functioning have been perfectly characterized. Such is not often the case, as  atmospheric conditions turn to strongly vary in time unpredictably, a fact that ultimately affects the transmission performance~\cite{Dequal2020,Andrews2005,Usenko2012a,Pirandola2021,Vasylyev2011,Vasylyev2017, bilkis2021reinforcement}.

We now revisit the re-calibration framework introduced in Sec.~\ref{sec:framework} for the Kennedy receiver. As stated above, the score function $\sct$ is given by the success probability of the communication protocol, which depends on the displacement value $\theta$, and the guessing rule $\hat{k}(\theta,n)$. Note that if access to outcome probabilities is granted, then the agent would perform a maximum-likelihood guess. However, in situations where such probabilities are not available, \textit{e.g.} no model of transmission channel, then the agent needs also to learn the optimal guessing rule. Thus, we remark that the score function is dependent on the specific transmission channel acting between sender and receiver, and potentially differs from the \textit{noise-less} success probability, \textit{i.e.} identity channel acting in between parties. The latter quantity provides the effective score $\sca$. Here, the intensity $|\alpha|^2$ is initially estimated using an $N_{\text{eff}}$ experiment repetitions, where the displacement value $\theta$ is set to zero, and thus the outcomes probabilities can be linked to $\alpha$ as per the Born rule, \textit{i.e.} $p(n=0|(-1)^k\alpha) = e^{-|(-1)^k\alpha|^2}$, with $\sum_{i=0,1}p(n=i|(-1)^k\alpha) = 1$. Thus, estimates of outcome probabilities are used to estimate the signal intensity, which is in turn used to initialize the state-action value functions $\llaves{Q(\theta)$, $Q(\hat{k};n,\theta)}$ to the success probability of setting displacement $\theta$ and conditional probabilities of having $\hat{k}$ given observation $n$ and displacement $\theta$ respectively.

The aforementioned quantities are then used by a Q-learning agent, which fine-tunes the calibrating strategy to the experimental conditions at hand; this is done by providing a binary reward to the agent according to the correctness of its guess $\hat{k}$, and it can be proven that such scheme converges to the optimal device configuration~\cite{bilkis2020real}. The Q-learning method is applied for $N_{rl}$ experiments, and then the receiver is \textit{deployed}. While in deployment stage, the agent monitors the measurement outcome statistics, by keeping track of their estimate values according to the past experiments. This quantity serves as a de-calibration witness $\Wd$, and abrupt changes of this quantity indicates that a change-point has occurred. If exiting an indecision region (specified by the agent), then the calibration protocol is re-started. 

\begin{figure}[t!]
    \centering
    \includegraphics[scale=0.27]{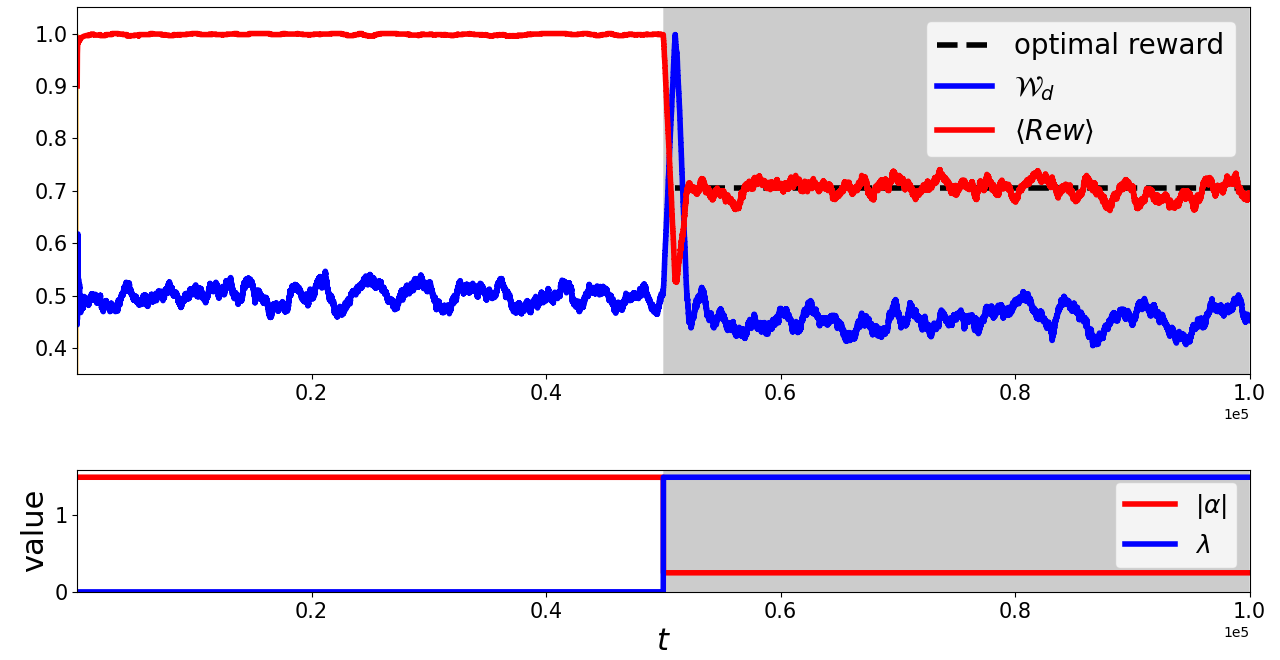}\\
    \includegraphics[scale=.27]{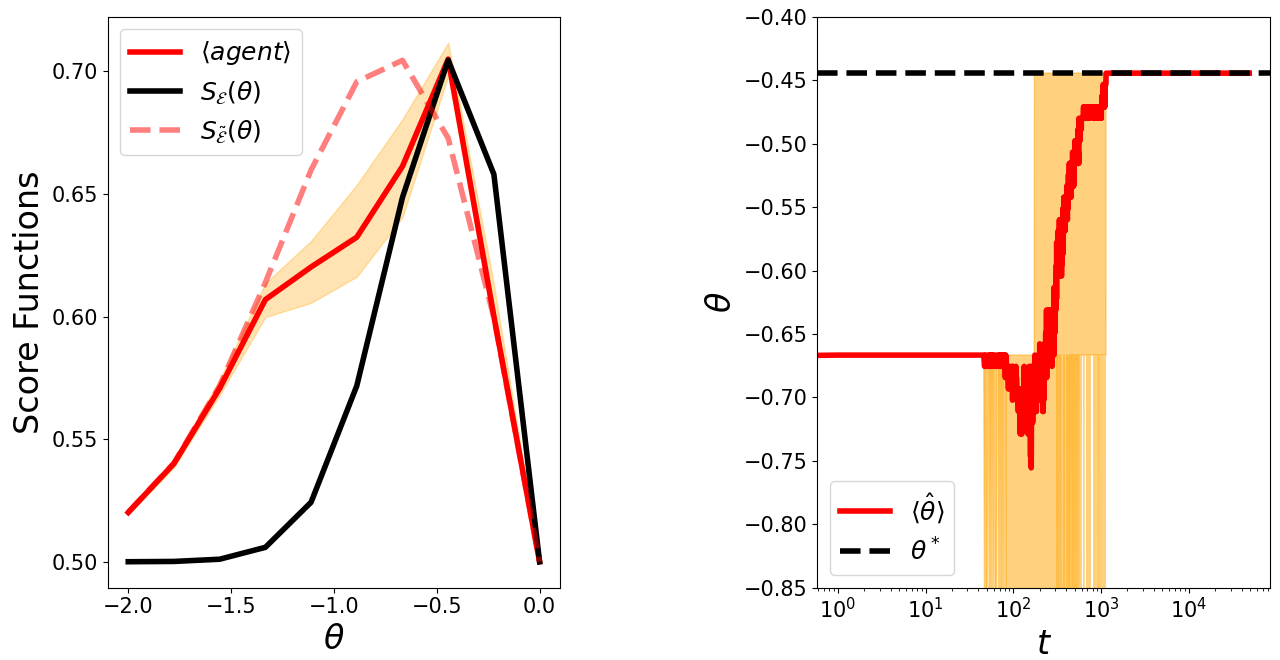}

    \caption{Recalibration and learning curve. \textit{(Top)}: Learning curve evolution. Average reward acquired during the last $10^3$ experiments $\langle Rew\rangle = \sum_{k=0}^{10^3}\frac{r_{t-k}}{10^3}$, and the evolution of the de-calibration witness $\Wd$ estimated by measurement statistics. As can be seen in the change-point, the Witness presents a big fluctuation, starting a recalibration of the system until the agent converges to the optimal reward.
    \textit{(Bottom)}: Update of the Q-values curve (left) and evolution of agent's \textit{greedy} strategy, \textit{i.e.} the configuration the calibrating agent would choose at each experiment (right).
    }
    \label{fig:witness_noise_and_beta_change}
\end{figure}

\textit{Malfunctioning device example}. We now consider the case in which the Kennedy receiver is initially calibrated to its optimal configuration, with pre-defined intensity value $|\alpha_0|^2$, deployed to ideal conditions for such initial environment $\mathcal{E}_0$, and incurs into a de-calibration. The new environment $\mathcal{E}_1$ consists of a different intensity value $|\alpha_1|^2$ of the signals arriving to the receiver, plus a \textit{faulty} displacement. Here, the value $\theta$ the agent fixes, actually displaces the signal by a value $\lambda \theta$, with $\lambda \geq 1$ being an unknown parameter. The effect of this faulty behavior is to make displacements bigger than expected, shifting the value of the optimal configuration $\theta^*$. As a consequence, the score-function landscape gets modified, in this case narrower. We remark that the malfunctioning behavior is unknown to the agent, who first loads the Q-values for the ideal case using its effective model $\sca$, and then fine-tunes them by Q-learning; specific details on the implementation are provided in Appendix~\ref{chap:append}.  

In Fig.~\ref{fig:witness_noise_and_beta_change} (top) we show the learning-curve of the agent in terms of cumulative reward acquired; we remark that since initially the effective model leads to a value that coincides with the score function of the device, the reward acquired is maximum (after an initial estimation round, which in this case consists on $10^3$ experiments, and is used to estimate $|\alpha|^2$ by setting the displacement $\theta = 0$). The de-calibration witness $\Wd$ is taken to be an estimate of the outcome probability $\hat{p}(n=1)$, and by monitoring abrupt changes in this quantity, the agent is able to detect environment shift $\mathcal{E}_0\rightarrow\mathcal{E}_1$. The change point in which the device enters a malfunctioning stage occurs at experiment $5\;10^5$, and can readily be seen in the top panel of Fig.~\ref{fig:witness_noise_and_beta_change} by the change of $\Wd$ behavior. Additionally, this can be detected by an abrupt change in the cumulative reward acquired; however, note that such quantity is potentially not available during deployment stage. As a consequence, the agent uses its change-point detection strategy to re-activate the calibration protocol again, by estimating the new signal intensity and initializing the Q-values in the effective model obtained thereby. Note that in this new scenario, the effective model does not coincide anymore with the underlying true one. This fact is illustrated by the initial and final Q-values obtained by the agent, shown in Fig.~\ref{fig:witness_noise_and_beta_change} (bottom), where we additionally show the optimal configuration $\hat{\theta}^*$ suggested by the agent at a given experiment.

\section{Outlook \& future research directions}\label{sec:chau}
In this work, we presented a re-calibration framework, accompanied by an automatic re-calibration method, and targeted to real-world quantum technology devices. 

We illustrated the proposed method using a Kennedy receiver under heavily varying deployment conditions, as an  example used to develop the proposed mathematical apparatus. As in any device, de-calibration is a frequent problem that needs to be addressed. This example serves as a test-bed for our automatic re-calibration framework, showing that not only the calibrating agent is able to configure the device in a semi-agnostic way, but also to detect situations in which the device gets off-calibrated. Our mechanism allows for the automation of the re-calibration process and is readily available to be implemented in a wide scope of technologies, even beyond quantum technology applications.

Specifically, our technique reduces the number of experiment repetitions needed to (re)-calibrate the device. This is done by making use of an effective model, whose purpose is to capture main features in the configuration landscape, and is complemented by model-free reinforcement-learning techniques. Additionally, we introduce the de-calibration witness statistic, which plays a key role in detecting either novelties or anomalies referring to device's functioning. Such quantity is conceived as a figure of merit to be calculated during device deployment. In this stage, the score function for the quality of the controls that are chosen by the agent is not readily computable, and the agent can only rely on information available in the experiment, \textit{e.g.} statistics from the measurement outcome . 

A plethora of change-point/anomaly detection methods can be used in order to complement our method~\cite{Page1955test,brodsky2010non,Basseville1993detection,Sentis2017exact,Fanizza2023ultimate,Sentis2016cgp}. However, let us remark that an alternative to monitoring the de-calibration witnesses can also be brought to attention, \textit{i.e} by presetting a calibration control routine. Such scheme demands balancing between device deploying and guaranteeing that the optimal configuration is being kept, and can potentially be implemented by allowing intermediate calibration stages in between deployment. Here, we remark that while model-free RL techniques could potentially adapt the controls to \textit{smooth} changes in the optimal configuration (without the necessity of an effective model nor a de-calibration witness), abrupt changes would in practice corrupt a successful adaptation. Here, an \textit{abruptness} notion is unveiled when it comes to environment changes: on the one hand we identify continual reinforcement learning ~\cite{abel2023definition,khetarpal2020towards} (where the calibration agent smoothly adapts the configuration as the environment \textit{smoothly} varies), and on the other hand domain adaptation in reinforcement learning~\cite{tobin2017domain,xing2021domain} (where the calibration needs to be adapted under changes of abrupt nature, as the ones considered in this paper). The setting studied in our paper might also be tackled from an active learning framework~\cite{Dadeva2012active}, in which the agent may inject prior knowledge on the different conditions in which the device is expected to be deployed, and can potentially be used to further exploit the symbiosis between model-free and model-aware routines considered above.

A straightforward extension of settings where our re-calibration framework finds real-world implementations is given by Noisy Intermediate-Scale Quantum (NISQ) devices, where the strong presence of noise severely limits the scope of applications, and developing tools to address such issues is an active area of research.

Furthermore, our work opens the door for several follow-up implementations and enhancements of the re-calibration protocol. Among them, usage of more sophisticated RL methods~\cite{Sutton2018, Baum2021Experimental} and inspecting the possibility of a coherent re-calibration by usage of quantum correlations~\cite{Liao2021quantum,abbas2023quantumopt}.

\vspace{.05cm}

\section{Acknowledgments} 
M.B. is grateful for useful discussions with John Calsamiglia about enhancing the Q-values initialization. T.C wants to thank F. Tomás B. Perez, for ideas in possible applications and numerical results. T.C. acknowledges the support from the CVC fellowship program. M.B. acknowledges support from AGAUR Grant no. 2023 INV-2 00034 funded by he European Union, Next Generation EU and Grant PID2021-126808OB-I00 funded by MCIN/AEI/ 10.13039/501100011033 and by ERDF A way of making Europe. F.V and M.B acknowledge the support from the Spanish Ministry of Science and Innovation through the project GRAIL, grant no. PID291-1268080B-100. J. M. M. and L.R. acknowledge to Consejo Nacional de Investigaciones Cient\'ificas y T\'ecnicas (CONICET) and support from ANPCyT Argentina, Pr\'estamo BID, Grant no. PICT 20203490.  

\bibliographystyle{apsrev4-1}
\bibliography{library}

\begin{thebibliography}{95}%
\makeatletter
\providecommand \@ifxundefined [1]{%
 \@ifx{#1\undefined}
}%
\providecommand \@ifnum [1]{%
 \ifnum #1\expandafter \@firstoftwo
 \else \expandafter \@secondoftwo
 \fi
}%
\providecommand \@ifx [1]{%
 \ifx #1\expandafter \@firstoftwo
 \else \expandafter \@secondoftwo
 \fi
}%
\providecommand \natexlab [1]{#1}%
\providecommand \enquote  [1]{``#1''}%
\providecommand \bibnamefont  [1]{#1}%
\providecommand \bibfnamefont [1]{#1}%
\providecommand \citenamefont [1]{#1}%
\providecommand \href@noop [0]{\@secondoftwo}%
\providecommand \href [0]{\begingroup \@sanitize@url \@href}%
\providecommand \@href[1]{\@@startlink{#1}\@@href}%
\providecommand \@@href[1]{\endgroup#1\@@endlink}%
\providecommand \@sanitize@url [0]{\catcode `\\12\catcode `\$12\catcode `\&12\catcode `\#12\catcode `\^12\catcode `\_12\catcode `\%12\relax}%
\providecommand \@@startlink[1]{}%
\providecommand \@@endlink[0]{}%
\providecommand \url  [0]{\begingroup\@sanitize@url \@url }%
\providecommand \@url [1]{\endgroup\@href {#1}{\urlprefix }}%
\providecommand \urlprefix  [0]{URL }%
\providecommand \Eprint [0]{\href }%
\providecommand \doibase [0]{http://dx.doi.org/}%
\providecommand \selectlanguage [0]{\@gobble}%
\providecommand \bibinfo  [0]{\@secondoftwo}%
\providecommand \bibfield  [0]{\@secondoftwo}%
\providecommand \translation [1]{[#1]}%
\providecommand \BibitemOpen [0]{}%
\providecommand \bibitemStop [0]{}%
\providecommand \bibitemNoStop [0]{.\EOS\space}%
\providecommand \EOS [0]{\spacefactor3000\relax}%
\providecommand \BibitemShut  [1]{\csname bibitem#1\endcsname}%
\let\auto@bib@innerbib\@empty
%</preamble>
\bibitem [{\citenamefont {Cimini}\ \emph {et~al.}(2021)\citenamefont {Cimini}, \citenamefont {Polino}, \citenamefont {Valeri}, \citenamefont {Gianani}, \citenamefont {Spagnolo}, \citenamefont {Corrielli}, \citenamefont {Crespi}, \citenamefont {Osellame}, \citenamefont {Barbieri},\ and\ \citenamefont {Sciarrino}}]{cimini2021calibration}%
  \BibitemOpen
  \bibfield  {author} {\bibinfo {author} {\bibfnamefont {V.}~\bibnamefont {Cimini}}, \bibinfo {author} {\bibfnamefont {E.}~\bibnamefont {Polino}}, \bibinfo {author} {\bibfnamefont {M.}~\bibnamefont {Valeri}}, \bibinfo {author} {\bibfnamefont {I.}~\bibnamefont {Gianani}}, \bibinfo {author} {\bibfnamefont {N.}~\bibnamefont {Spagnolo}}, \bibinfo {author} {\bibfnamefont {G.}~\bibnamefont {Corrielli}}, \bibinfo {author} {\bibfnamefont {A.}~\bibnamefont {Crespi}}, \bibinfo {author} {\bibfnamefont {R.}~\bibnamefont {Osellame}}, \bibinfo {author} {\bibfnamefont {M.}~\bibnamefont {Barbieri}}, \ and\ \bibinfo {author} {\bibfnamefont {F.}~\bibnamefont {Sciarrino}},\ }\href {\doibase 10.1103/physrevapplied.15.044003} {\bibfield  {journal} {\bibinfo  {journal} {Physical Review Applied}\ }\textbf {\bibinfo {volume} {15}} (\bibinfo {year} {2021}),\ 10.1103/physrevapplied.15.044003}\BibitemShut {NoStop}%
\bibitem [{\citenamefont {Cimini}\ \emph {et~al.}(2024)\citenamefont {Cimini}, \citenamefont {Valeri}, \citenamefont {Piacentini}, \citenamefont {Ceccarelli}, \citenamefont {Corrielli}, \citenamefont {Osellame}, \citenamefont {Spagnolo},\ and\ \citenamefont {Sciarrino}}]{cimini2024variational}%
  \BibitemOpen
  \bibfield  {author} {\bibinfo {author} {\bibfnamefont {V.}~\bibnamefont {Cimini}}, \bibinfo {author} {\bibfnamefont {M.}~\bibnamefont {Valeri}}, \bibinfo {author} {\bibfnamefont {S.}~\bibnamefont {Piacentini}}, \bibinfo {author} {\bibfnamefont {F.}~\bibnamefont {Ceccarelli}}, \bibinfo {author} {\bibfnamefont {G.}~\bibnamefont {Corrielli}}, \bibinfo {author} {\bibfnamefont {R.}~\bibnamefont {Osellame}}, \bibinfo {author} {\bibfnamefont {N.}~\bibnamefont {Spagnolo}}, \ and\ \bibinfo {author} {\bibfnamefont {F.}~\bibnamefont {Sciarrino}},\ }\href {\doibase 10.1038/s41534-024-00821-0} {\bibfield  {journal} {\bibinfo  {journal} {npj Quantum Information}\ }\textbf {\bibinfo {volume} {10}} (\bibinfo {year} {2024}),\ 10.1038/s41534-024-00821-0}\BibitemShut {NoStop}%
\bibitem [{\citenamefont {Ren}\ \emph {et~al.}(2020)\citenamefont {Ren}, \citenamefont {Yang}, \citenamefont {Liu}, \citenamefont {Huang},\ and\ \citenamefont {Guo}}]{ren2020sensor}%
  \BibitemOpen
  \bibfield  {author} {\bibinfo {author} {\bibfnamefont {H.}~\bibnamefont {Ren}}, \bibinfo {author} {\bibfnamefont {J.}~\bibnamefont {Yang}}, \bibinfo {author} {\bibfnamefont {X.}~\bibnamefont {Liu}}, \bibinfo {author} {\bibfnamefont {P.}~\bibnamefont {Huang}}, \ and\ \bibinfo {author} {\bibfnamefont {L.}~\bibnamefont {Guo}},\ }\href {\doibase 10.3390/s20133779} {\bibfield  {journal} {\bibinfo  {journal} {Sensors}\ }\textbf {\bibinfo {volume} {20}} (\bibinfo {year} {2020}),\ 10.3390/s20133779}\BibitemShut {NoStop}%
\bibitem [{\citenamefont {Vernuccio}\ \emph {et~al.}(2022)\citenamefont {Vernuccio}, \citenamefont {Bresci}, \citenamefont {Cimini}, \citenamefont {Giuseppi}, \citenamefont {Cerullo}, \citenamefont {Polli},\ and\ \citenamefont {Valensise}}]{vernuccio2022artificial}%
  \BibitemOpen
  \bibfield  {author} {\bibinfo {author} {\bibfnamefont {F.}~\bibnamefont {Vernuccio}}, \bibinfo {author} {\bibfnamefont {A.}~\bibnamefont {Bresci}}, \bibinfo {author} {\bibfnamefont {V.}~\bibnamefont {Cimini}}, \bibinfo {author} {\bibfnamefont {A.}~\bibnamefont {Giuseppi}}, \bibinfo {author} {\bibfnamefont {G.}~\bibnamefont {Cerullo}}, \bibinfo {author} {\bibfnamefont {D.}~\bibnamefont {Polli}}, \ and\ \bibinfo {author} {\bibfnamefont {C.~M.}\ \bibnamefont {Valensise}},\ }\href {\doibase 10.1002/lpor.202100399} {\bibfield  {journal} {\bibinfo  {journal} {Laser \& Photonics Reviews}\ }\textbf {\bibinfo {volume} {16}} (\bibinfo {year} {2022}),\ 10.1002/lpor.202100399}\BibitemShut {NoStop}%
\bibitem [{\citenamefont {Ono}(2016)}]{ono2016calibration}%
  \BibitemOpen
  \bibfield  {author} {\bibinfo {author} {\bibfnamefont {K.}~\bibnamefont {Ono}},\ }\href {\doibase 10.3390/ma9070508} {\bibfield  {journal} {\bibinfo  {journal} {Materials}\ }\textbf {\bibinfo {volume} {9}} (\bibinfo {year} {2016}),\ 10.3390/ma9070508}\BibitemShut {NoStop}%
\bibitem [{\citenamefont {Zhao}\ \emph {et~al.}(2019)\citenamefont {Zhao}, \citenamefont {Liu},\ and\ \citenamefont {Li}}]{zhao2019online}%
  \BibitemOpen
  \bibfield  {author} {\bibinfo {author} {\bibfnamefont {S.}~\bibnamefont {Zhao}}, \bibinfo {author} {\bibfnamefont {J.}~\bibnamefont {Liu}}, \ and\ \bibinfo {author} {\bibfnamefont {Y.}~\bibnamefont {Li}},\ }\href {\doibase 10.3390/en12101923} {\bibfield  {journal} {\bibinfo  {journal} {Energies}\ }\textbf {\bibinfo {volume} {12}} (\bibinfo {year} {2019}),\ 10.3390/en12101923}\BibitemShut {NoStop}%
\bibitem [{\citenamefont {Haitjema}(2020)}]{haitjema2020the}%
  \BibitemOpen
  \bibfield  {author} {\bibinfo {author} {\bibfnamefont {H.}~\bibnamefont {Haitjema}},\ }\href {\doibase 10.3390/s20030584} {\bibfield  {journal} {\bibinfo  {journal} {Sensors}\ }\textbf {\bibinfo {volume} {20}} (\bibinfo {year} {2020}),\ 10.3390/s20030584}\BibitemShut {NoStop}%
\bibitem [{\citenamefont {Cimini}\ \emph {et~al.}(2019)\citenamefont {Cimini}, \citenamefont {Gianani}, \citenamefont {Spagnolo}, \citenamefont {Leccese}, \citenamefont {Sciarrino},\ and\ \citenamefont {Barbieri}}]{cimini2019calibration}%
  \BibitemOpen
  \bibfield  {author} {\bibinfo {author} {\bibfnamefont {V.}~\bibnamefont {Cimini}}, \bibinfo {author} {\bibfnamefont {I.}~\bibnamefont {Gianani}}, \bibinfo {author} {\bibfnamefont {N.}~\bibnamefont {Spagnolo}}, \bibinfo {author} {\bibfnamefont {F.}~\bibnamefont {Leccese}}, \bibinfo {author} {\bibfnamefont {F.}~\bibnamefont {Sciarrino}}, \ and\ \bibinfo {author} {\bibfnamefont {M.}~\bibnamefont {Barbieri}},\ }\href {\doibase 10.1103/physrevlett.123.230502} {\bibfield  {journal} {\bibinfo  {journal} {Physical Review Letters}\ }\textbf {\bibinfo {volume} {123}} (\bibinfo {year} {2019}),\ 10.1103/physrevlett.123.230502}\BibitemShut {NoStop}%
\bibitem [{\citenamefont {Zhang}\ \emph {et~al.}(2021)\citenamefont {Zhang}, \citenamefont {Wijeratne}, \citenamefont {Talebi},\ and\ \citenamefont {Lary}}]{zhang2021machine}%
  \BibitemOpen
  \bibfield  {author} {\bibinfo {author} {\bibfnamefont {Y.}~\bibnamefont {Zhang}}, \bibinfo {author} {\bibfnamefont {L.~O.~H.}\ \bibnamefont {Wijeratne}}, \bibinfo {author} {\bibfnamefont {S.}~\bibnamefont {Talebi}}, \ and\ \bibinfo {author} {\bibfnamefont {D.~J.}\ \bibnamefont {Lary}},\ }\href {\doibase 10.3390/s21186259} {\bibfield  {journal} {\bibinfo  {journal} {Sensors}\ }\textbf {\bibinfo {volume} {21}} (\bibinfo {year} {2021}),\ 10.3390/s21186259}\BibitemShut {NoStop}%
\bibitem [{\citenamefont {Bilkis}\ \emph {et~al.}(2020)\citenamefont {Bilkis}, \citenamefont {Rosati}, \citenamefont {Yepes},\ and\ \citenamefont {Calsamiglia}}]{bilkis2020real}%
  \BibitemOpen
  \bibfield  {author} {\bibinfo {author} {\bibfnamefont {M.}~\bibnamefont {Bilkis}}, \bibinfo {author} {\bibfnamefont {M.}~\bibnamefont {Rosati}}, \bibinfo {author} {\bibfnamefont {R.~M.}\ \bibnamefont {Yepes}}, \ and\ \bibinfo {author} {\bibfnamefont {J.}~\bibnamefont {Calsamiglia}},\ }\href {\doibase 10.1103/PhysRevResearch.2.033295} {\bibfield  {journal} {\bibinfo  {journal} {Phys. Rev. Res.}\ }\textbf {\bibinfo {volume} {2}},\ \bibinfo {pages} {033295} (\bibinfo {year} {2020})}\BibitemShut {NoStop}%
\bibitem [{\citenamefont {Bilkis}\ \emph {et~al.}(2021)\citenamefont {Bilkis}, \citenamefont {Rosati},\ and\ \citenamefont {Calsamiglia}}]{bilkis2021reinforcement}%
  \BibitemOpen
  \bibfield  {author} {\bibinfo {author} {\bibfnamefont {M.}~\bibnamefont {Bilkis}}, \bibinfo {author} {\bibfnamefont {M.}~\bibnamefont {Rosati}}, \ and\ \bibinfo {author} {\bibfnamefont {J.}~\bibnamefont {Calsamiglia}},\ }in\ \href {\doibase 10.1109/ITW48936.2021.9611396} {\emph {\bibinfo {booktitle} {2021 IEEE Information Theory Workshop (ITW)}}}\ (\bibinfo {year} {2021})\ pp.\ \bibinfo {pages} {1--6}\BibitemShut {NoStop}%
\bibitem [{\citenamefont {Dequal}\ \emph {et~al.}(2019)\citenamefont {Dequal}, \citenamefont {Vidarte}, \citenamefont {Rodriguez}, \citenamefont {Leverrier}, \citenamefont {Vallone}, \citenamefont {Villoresi},\ and\ \citenamefont {Diamanti}}]{Dequal2020}%
  \BibitemOpen
  \bibfield  {author} {\bibinfo {author} {\bibfnamefont {D.}~\bibnamefont {Dequal}}, \bibinfo {author} {\bibfnamefont {L.~T.}\ \bibnamefont {Vidarte}}, \bibinfo {author} {\bibfnamefont {V.~R.}\ \bibnamefont {Rodriguez}}, \bibinfo {author} {\bibfnamefont {A.}~\bibnamefont {Leverrier}}, \bibinfo {author} {\bibfnamefont {G.}~\bibnamefont {Vallone}}, \bibinfo {author} {\bibfnamefont {P.}~\bibnamefont {Villoresi}}, \ and\ \bibinfo {author} {\bibfnamefont {E.}~\bibnamefont {Diamanti}},\ }in\ \href {\doibase 10.1364/QIM.2019.T5A.89} {\emph {\bibinfo {booktitle} {Quantum Inf. Meas. V Quantum Technol.}}},\ Vol.\ \bibinfo {volume} {Part F165-}\ (\bibinfo  {publisher} {OSA},\ \bibinfo {address} {Washington, D.C.},\ \bibinfo {year} {2019})\ p.\ \bibinfo {pages} {T5A.89},\ \Eprint {http://arxiv.org/abs/2002.02002} {arXiv:2002.02002} \BibitemShut {NoStop}%
\bibitem [{\citenamefont {Andrews}\ and\ \citenamefont {Phillips}(2005)}]{Andrews2005}%
  \BibitemOpen
  \bibfield  {author} {\bibinfo {author} {\bibfnamefont {L.~C.}\ \bibnamefont {Andrews}}\ and\ \bibinfo {author} {\bibfnamefont {R.~L.}\ \bibnamefont {Phillips}},\ }\href {\doibase 10.1117/3.626196} {\emph {\bibinfo {title} {{Laser Beam Propagation through Random Media}}}}\ (\bibinfo  {publisher} {SPIE},\ \bibinfo {address} {1000 20th Street, Bellingham, WA 98227-0010 USA},\ \bibinfo {year} {2005})\BibitemShut {NoStop}%
\bibitem [{\citenamefont {Pirandola}(2021{\natexlab{a}})}]{Pirandola2021a}%
  \BibitemOpen
  \bibfield  {author} {\bibinfo {author} {\bibfnamefont {S.}~\bibnamefont {Pirandola}},\ }\href {\doibase 10.1103/PhysRevResearch.3.023130} {\bibfield  {journal} {\bibinfo  {journal} {Phys. Rev. Res.}\ }\textbf {\bibinfo {volume} {3}},\ \bibinfo {pages} {023130} (\bibinfo {year} {2021}{\natexlab{a}})}\BibitemShut {NoStop}%
\bibitem [{\citenamefont {Pirandola}(2021{\natexlab{b}})}]{Pirandola2021}%
  \BibitemOpen
  \bibfield  {author} {\bibinfo {author} {\bibfnamefont {S.}~\bibnamefont {Pirandola}},\ }\href {\doibase 10.1103/PhysRevResearch.3.013279} {\bibfield  {journal} {\bibinfo  {journal} {Phys. Rev. Res.}\ }\textbf {\bibinfo {volume} {3}},\ \bibinfo {pages} {013279} (\bibinfo {year} {2021}{\natexlab{b}})}\BibitemShut {NoStop}%
\bibitem [{\citenamefont {Vasylyev}\ \emph {et~al.}(2012)\citenamefont {Vasylyev}, \citenamefont {Semenov},\ and\ \citenamefont {Vogel}}]{Vasylyev2011}%
  \BibitemOpen
  \bibfield  {author} {\bibinfo {author} {\bibfnamefont {D.~Y.}\ \bibnamefont {Vasylyev}}, \bibinfo {author} {\bibfnamefont {A.~A.}\ \bibnamefont {Semenov}}, \ and\ \bibinfo {author} {\bibfnamefont {W.}~\bibnamefont {Vogel}},\ }\href {\doibase 10.1103/PhysRevLett.108.220501} {\bibfield  {journal} {\bibinfo  {journal} {Phys. Rev. Lett.}\ }\textbf {\bibinfo {volume} {108}},\ \bibinfo {pages} {220501} (\bibinfo {year} {2012})}\BibitemShut {NoStop}%
\bibitem [{\citenamefont {Vasylyev}\ \emph {et~al.}(2017)\citenamefont {Vasylyev}, \citenamefont {Semenov}, \citenamefont {Vogel}, \citenamefont {G\"unthner}, \citenamefont {Thurn}, \citenamefont {Bayraktar},\ and\ \citenamefont {Marquardt}}]{Vasylyev2017}%
  \BibitemOpen
  \bibfield  {author} {\bibinfo {author} {\bibfnamefont {D.}~\bibnamefont {Vasylyev}}, \bibinfo {author} {\bibfnamefont {A.~A.}\ \bibnamefont {Semenov}}, \bibinfo {author} {\bibfnamefont {W.}~\bibnamefont {Vogel}}, \bibinfo {author} {\bibfnamefont {K.}~\bibnamefont {G\"unthner}}, \bibinfo {author} {\bibfnamefont {A.}~\bibnamefont {Thurn}}, \bibinfo {author} {\bibfnamefont {O.}~\bibnamefont {Bayraktar}}, \ and\ \bibinfo {author} {\bibfnamefont {C.}~\bibnamefont {Marquardt}},\ }\href {\doibase 10.1103/PhysRevA.96.043856} {\bibfield  {journal} {\bibinfo  {journal} {Phys. Rev. A}\ }\textbf {\bibinfo {volume} {96}},\ \bibinfo {pages} {043856} (\bibinfo {year} {2017})}\BibitemShut {NoStop}%
\bibitem [{\citenamefont {Jung}\ \emph {et~al.}(2017)\citenamefont {Jung}, \citenamefont {Choi},\ and\ \citenamefont {Kim}}]{jung2017multiobjective}%
  \BibitemOpen
  \bibfield  {author} {\bibinfo {author} {\bibfnamefont {D.}~\bibnamefont {Jung}}, \bibinfo {author} {\bibfnamefont {Y.}~\bibnamefont {Choi}}, \ and\ \bibinfo {author} {\bibfnamefont {J.}~\bibnamefont {Kim}},\ }\href {\doibase 10.3390/w9030187} {\bibfield  {journal} {\bibinfo  {journal} {Water}\ }\textbf {\bibinfo {volume} {9}} (\bibinfo {year} {2017}),\ 10.3390/w9030187}\BibitemShut {NoStop}%
\bibitem [{\citenamefont {Kavetski}\ \emph {et~al.}(2006)\citenamefont {Kavetski}, \citenamefont {Kuczera},\ and\ \citenamefont {Franks}}]{KAVETSKI2006173}%
  \BibitemOpen
  \bibfield  {author} {\bibinfo {author} {\bibfnamefont {D.}~\bibnamefont {Kavetski}}, \bibinfo {author} {\bibfnamefont {G.}~\bibnamefont {Kuczera}}, \ and\ \bibinfo {author} {\bibfnamefont {S.~W.}\ \bibnamefont {Franks}},\ }\href {\doibase https://doi.org/10.1016/j.jhydrol.2005.07.012} {\bibfield  {journal} {\bibinfo  {journal} {Journal of Hydrology}\ }\textbf {\bibinfo {volume} {320}},\ \bibinfo {pages} {173} (\bibinfo {year} {2006})},\ \bibinfo {note} {the model parameter estimation experiment}\BibitemShut {NoStop}%
\bibitem [{\citenamefont {Papafotis}\ \emph {et~al.}(2021)\citenamefont {Papafotis}, \citenamefont {Nikitas},\ and\ \citenamefont {Sotiriadis}}]{papafotis2021magnetic}%
  \BibitemOpen
  \bibfield  {author} {\bibinfo {author} {\bibfnamefont {K.}~\bibnamefont {Papafotis}}, \bibinfo {author} {\bibfnamefont {D.}~\bibnamefont {Nikitas}}, \ and\ \bibinfo {author} {\bibfnamefont {P.~P.}\ \bibnamefont {Sotiriadis}},\ }\href {\doibase 10.3390/s21165288} {\bibfield  {journal} {\bibinfo  {journal} {Sensors}\ }\textbf {\bibinfo {volume} {21}} (\bibinfo {year} {2021}),\ 10.3390/s21165288}\BibitemShut {NoStop}%
\bibitem [{\citenamefont {Cao}\ \emph {et~al.}(2020)\citenamefont {Cao}, \citenamefont {Xu},\ and\ \citenamefont {Xu}}]{cao2020realtime}%
  \BibitemOpen
  \bibfield  {author} {\bibinfo {author} {\bibfnamefont {G.}~\bibnamefont {Cao}}, \bibinfo {author} {\bibfnamefont {X.}~\bibnamefont {Xu}}, \ and\ \bibinfo {author} {\bibfnamefont {D.}~\bibnamefont {Xu}},\ }\href {\doibase 10.3390/s20020535} {\bibfield  {journal} {\bibinfo  {journal} {Sensors}\ }\textbf {\bibinfo {volume} {20}} (\bibinfo {year} {2020}),\ 10.3390/s20020535}\BibitemShut {NoStop}%
\bibitem [{\citenamefont {Fallani}\ \emph {et~al.}(2022)\citenamefont {Fallani}, \citenamefont {Rossi}, \citenamefont {Tamascelli},\ and\ \citenamefont {Genoni}}]{fallani2022learning}%
  \BibitemOpen
  \bibfield  {author} {\bibinfo {author} {\bibfnamefont {A.}~\bibnamefont {Fallani}}, \bibinfo {author} {\bibfnamefont {M.~A.~C.}\ \bibnamefont {Rossi}}, \bibinfo {author} {\bibfnamefont {D.}~\bibnamefont {Tamascelli}}, \ and\ \bibinfo {author} {\bibfnamefont {M.~G.}\ \bibnamefont {Genoni}},\ }\href {\doibase 10.1103/PRXQuantum.3.020310} {\bibfield  {journal} {\bibinfo  {journal} {PRX Quantum}\ }\textbf {\bibinfo {volume} {3}},\ \bibinfo {pages} {020310} (\bibinfo {year} {2022})}\BibitemShut {NoStop}%
\bibitem [{\citenamefont {Fiderer}\ \emph {et~al.}(2021)\citenamefont {Fiderer}, \citenamefont {Schuff},\ and\ \citenamefont {Braun}}]{fiderer2021neural}%
  \BibitemOpen
  \bibfield  {author} {\bibinfo {author} {\bibfnamefont {L.~J.}\ \bibnamefont {Fiderer}}, \bibinfo {author} {\bibfnamefont {J.}~\bibnamefont {Schuff}}, \ and\ \bibinfo {author} {\bibfnamefont {D.}~\bibnamefont {Braun}},\ }\href {\doibase 10.1103/PRXQuantum.2.020303} {\bibfield  {journal} {\bibinfo  {journal} {PRX Quantum}\ }\textbf {\bibinfo {volume} {2}},\ \bibinfo {pages} {020303} (\bibinfo {year} {2021})}\BibitemShut {NoStop}%
\bibitem [{\citenamefont {Fiderer}\ and\ \citenamefont {Braun}(2018)}]{Fiderer2018qchaos}%
  \BibitemOpen
  \bibfield  {author} {\bibinfo {author} {\bibfnamefont {L.~J.}\ \bibnamefont {Fiderer}}\ and\ \bibinfo {author} {\bibfnamefont {D.}~\bibnamefont {Braun}},\ }\href {\doibase 10.1038/s41467-018-03623-z} {\bibfield  {journal} {\bibinfo  {journal} {Nature Communications}\ }\textbf {\bibinfo {volume} {9}},\ \bibinfo {pages} {1351} (\bibinfo {year} {2018})}\BibitemShut {NoStop}%
\bibitem [{\citenamefont {Lee}\ \emph {et~al.}(2021)\citenamefont {Lee}, \citenamefont {Lawrie}, \citenamefont {Pooser}, \citenamefont {Lee}, \citenamefont {Rockstuhl},\ and\ \citenamefont {Tame}}]{lee2021quantum}%
  \BibitemOpen
  \bibfield  {author} {\bibinfo {author} {\bibfnamefont {C.}~\bibnamefont {Lee}}, \bibinfo {author} {\bibfnamefont {B.}~\bibnamefont {Lawrie}}, \bibinfo {author} {\bibfnamefont {R.}~\bibnamefont {Pooser}}, \bibinfo {author} {\bibfnamefont {K.-G.}\ \bibnamefont {Lee}}, \bibinfo {author} {\bibfnamefont {C.}~\bibnamefont {Rockstuhl}}, \ and\ \bibinfo {author} {\bibfnamefont {M.}~\bibnamefont {Tame}},\ }\href {\doibase 10.1021/acs.chemrev.0c01028} {\bibfield  {journal} {\bibinfo  {journal} {Chemical Reviews}\ }\textbf {\bibinfo {volume} {121}} (\bibinfo {year} {2021}),\ 10.1021/acs.chemrev.0c01028}\BibitemShut {NoStop}%
\bibitem [{\citenamefont {Nolan}\ \emph {et~al.}(2021)\citenamefont {Nolan}, \citenamefont {Smerzi},\ and\ \citenamefont {Pezzè}}]{nolan2021a}%
  \BibitemOpen
  \bibfield  {author} {\bibinfo {author} {\bibfnamefont {S.}~\bibnamefont {Nolan}}, \bibinfo {author} {\bibfnamefont {A.}~\bibnamefont {Smerzi}}, \ and\ \bibinfo {author} {\bibfnamefont {L.}~\bibnamefont {Pezzè}},\ }\href {\doibase 10.1038/s41534-021-00497-w} {\bibfield  {journal} {\bibinfo  {journal} {npj Quantum Information}\ }\textbf {\bibinfo {volume} {7}} (\bibinfo {year} {2021}),\ 10.1038/s41534-021-00497-w}\BibitemShut {NoStop}%
\bibitem [{\citenamefont {Ban}\ \emph {et~al.}(2021)\citenamefont {Ban}, \citenamefont {Echanobe}, \citenamefont {Ding}, \citenamefont {Puebla},\ and\ \citenamefont {Casanova}}]{ban2021neuralnetworkbased}%
  \BibitemOpen
  \bibfield  {author} {\bibinfo {author} {\bibfnamefont {Y.}~\bibnamefont {Ban}}, \bibinfo {author} {\bibfnamefont {J.}~\bibnamefont {Echanobe}}, \bibinfo {author} {\bibfnamefont {Y.}~\bibnamefont {Ding}}, \bibinfo {author} {\bibfnamefont {R.}~\bibnamefont {Puebla}}, \ and\ \bibinfo {author} {\bibfnamefont {J.}~\bibnamefont {Casanova}},\ }\href {\doibase 10.1088/2058-9565/ac16ed} {\bibfield  {journal} {\bibinfo  {journal} {Quantum Science and Technology}\ }\textbf {\bibinfo {volume} {6}} (\bibinfo {year} {2021}),\ 10.1088/2058-9565/ac16ed}\BibitemShut {NoStop}%
\bibitem [{\citenamefont {Chen}\ \emph {et~al.}(2022)\citenamefont {Chen}, \citenamefont {Ban}, \citenamefont {He}, \citenamefont {Cui}, \citenamefont {Huang}, \citenamefont {Li}, \citenamefont {Guo},\ and\ \citenamefont {Casanova}}]{chen2022a}%
  \BibitemOpen
  \bibfield  {author} {\bibinfo {author} {\bibfnamefont {Y.}~\bibnamefont {Chen}}, \bibinfo {author} {\bibfnamefont {Y.}~\bibnamefont {Ban}}, \bibinfo {author} {\bibfnamefont {R.}~\bibnamefont {He}}, \bibinfo {author} {\bibfnamefont {J.-M.}\ \bibnamefont {Cui}}, \bibinfo {author} {\bibfnamefont {Y.-F.}\ \bibnamefont {Huang}}, \bibinfo {author} {\bibfnamefont {C.-F.}\ \bibnamefont {Li}}, \bibinfo {author} {\bibfnamefont {G.-C.}\ \bibnamefont {Guo}}, \ and\ \bibinfo {author} {\bibfnamefont {J.}~\bibnamefont {Casanova}},\ }\href {\doibase 10.1038/s41534-022-00669-2} {\bibfield  {journal} {\bibinfo  {journal} {npj Quantum Information}\ }\textbf {\bibinfo {volume} {8}} (\bibinfo {year} {2022}),\ 10.1038/s41534-022-00669-2}\BibitemShut {NoStop}%
\bibitem [{\citenamefont {Rambhatla}\ \emph {et~al.}(2020)\citenamefont {Rambhatla}, \citenamefont {D'Aurelio}, \citenamefont {Valeri}, \citenamefont {Polino}, \citenamefont {Spagnolo},\ and\ \citenamefont {Sciarrino}}]{rambhatla2020adaptive}%
  \BibitemOpen
  \bibfield  {author} {\bibinfo {author} {\bibfnamefont {K.}~\bibnamefont {Rambhatla}}, \bibinfo {author} {\bibfnamefont {S.~E.}\ \bibnamefont {D'Aurelio}}, \bibinfo {author} {\bibfnamefont {M.}~\bibnamefont {Valeri}}, \bibinfo {author} {\bibfnamefont {E.}~\bibnamefont {Polino}}, \bibinfo {author} {\bibfnamefont {N.}~\bibnamefont {Spagnolo}}, \ and\ \bibinfo {author} {\bibfnamefont {F.}~\bibnamefont {Sciarrino}},\ }\href {\doibase 10.1103/physrevresearch.2.033078} {\bibfield  {journal} {\bibinfo  {journal} {Physical Review Research}\ }\textbf {\bibinfo {volume} {2}} (\bibinfo {year} {2020}),\ 10.1103/physrevresearch.2.033078}\BibitemShut {NoStop}%
\bibitem [{\citenamefont {Sent\'{\i}s}\ \emph {et~al.}(2016)\citenamefont {Sent\'{\i}s}, \citenamefont {Bagan}, \citenamefont {Calsamiglia}, \citenamefont {Chiribella},\ and\ \citenamefont {Mu\~noz Tapia}}]{Sentis2016cgp}%
  \BibitemOpen
  \bibfield  {author} {\bibinfo {author} {\bibfnamefont {G.}~\bibnamefont {Sent\'{\i}s}}, \bibinfo {author} {\bibfnamefont {E.}~\bibnamefont {Bagan}}, \bibinfo {author} {\bibfnamefont {J.}~\bibnamefont {Calsamiglia}}, \bibinfo {author} {\bibfnamefont {G.}~\bibnamefont {Chiribella}}, \ and\ \bibinfo {author} {\bibfnamefont {R.}~\bibnamefont {Mu\~noz Tapia}},\ }\href {\doibase 10.1103/PhysRevLett.117.150502} {\bibfield  {journal} {\bibinfo  {journal} {Phys. Rev. Lett.}\ }\textbf {\bibinfo {volume} {117}},\ \bibinfo {pages} {150502} (\bibinfo {year} {2016})}\BibitemShut {NoStop}%
\bibitem [{\citenamefont {Fanizza}\ \emph {et~al.}(2023)\citenamefont {Fanizza}, \citenamefont {Hirche},\ and\ \citenamefont {Calsamiglia}}]{Fanizza2023ultimate}%
  \BibitemOpen
  \bibfield  {author} {\bibinfo {author} {\bibfnamefont {M.}~\bibnamefont {Fanizza}}, \bibinfo {author} {\bibfnamefont {C.}~\bibnamefont {Hirche}}, \ and\ \bibinfo {author} {\bibfnamefont {J.}~\bibnamefont {Calsamiglia}},\ }\href {\doibase 10.1103/PhysRevLett.131.020602} {\bibfield  {journal} {\bibinfo  {journal} {Phys. Rev. Lett.}\ }\textbf {\bibinfo {volume} {131}},\ \bibinfo {pages} {020602} (\bibinfo {year} {2023})}\BibitemShut {NoStop}%
\bibitem [{\citenamefont {Sent\'{\i}s}\ \emph {et~al.}(2017)\citenamefont {Sent\'{\i}s}, \citenamefont {Calsamiglia},\ and\ \citenamefont {Mu\~noz Tapia}}]{Sentis2017exact}%
  \BibitemOpen
  \bibfield  {author} {\bibinfo {author} {\bibfnamefont {G.}~\bibnamefont {Sent\'{\i}s}}, \bibinfo {author} {\bibfnamefont {J.}~\bibnamefont {Calsamiglia}}, \ and\ \bibinfo {author} {\bibfnamefont {R.}~\bibnamefont {Mu\~noz Tapia}},\ }\href {\doibase 10.1103/PhysRevLett.119.140506} {\bibfield  {journal} {\bibinfo  {journal} {Phys. Rev. Lett.}\ }\textbf {\bibinfo {volume} {119}},\ \bibinfo {pages} {140506} (\bibinfo {year} {2017})}\BibitemShut {NoStop}%
\bibitem [{\citenamefont {Woźniak}\ \emph {et~al.}(2023)\citenamefont {Woźniak}, \citenamefont {Belis}, \citenamefont {Puljak}, \citenamefont {Barkoutsos}, \citenamefont {Dissertori}, \citenamefont {Grossi}, \citenamefont {Pierini}, \citenamefont {Reiter}, \citenamefont {Tavernelli},\ and\ \citenamefont {Vallecorsa}}]{Wozniak2023quantumanomaly}%
  \BibitemOpen
  \bibfield  {author} {\bibinfo {author} {\bibfnamefont {K.~A.}\ \bibnamefont {Woźniak}}, \bibinfo {author} {\bibfnamefont {V.}~\bibnamefont {Belis}}, \bibinfo {author} {\bibfnamefont {E.}~\bibnamefont {Puljak}}, \bibinfo {author} {\bibfnamefont {P.}~\bibnamefont {Barkoutsos}}, \bibinfo {author} {\bibfnamefont {G.}~\bibnamefont {Dissertori}}, \bibinfo {author} {\bibfnamefont {M.}~\bibnamefont {Grossi}}, \bibinfo {author} {\bibfnamefont {M.}~\bibnamefont {Pierini}}, \bibinfo {author} {\bibfnamefont {F.}~\bibnamefont {Reiter}}, \bibinfo {author} {\bibfnamefont {I.}~\bibnamefont {Tavernelli}}, \ and\ \bibinfo {author} {\bibfnamefont {S.}~\bibnamefont {Vallecorsa}},\ }\href@noop {} {\enquote {\bibinfo {title} {Quantum anomaly detection in the latent space of proton collision events at the lhc},}\ } (\bibinfo {year} {2023}),\ \Eprint {http://arxiv.org/abs/arXiv:2301.10780} {arXiv:2301.10780} \BibitemShut {NoStop}%
\bibitem [{\citenamefont {Baker}\ \emph {et~al.}(2022)\citenamefont {Baker}, \citenamefont {Horowitz}, \citenamefont {Radha}, \citenamefont {Fernandes}, \citenamefont {Jones}, \citenamefont {Noorani}, \citenamefont {Skavysh}, \citenamefont {Lamontangne},\ and\ \citenamefont {Sanders}}]{baker2022quantumvariational}%
  \BibitemOpen
  \bibfield  {author} {\bibinfo {author} {\bibfnamefont {J.~S.}\ \bibnamefont {Baker}}, \bibinfo {author} {\bibfnamefont {H.}~\bibnamefont {Horowitz}}, \bibinfo {author} {\bibfnamefont {S.~K.}\ \bibnamefont {Radha}}, \bibinfo {author} {\bibfnamefont {S.}~\bibnamefont {Fernandes}}, \bibinfo {author} {\bibfnamefont {C.}~\bibnamefont {Jones}}, \bibinfo {author} {\bibfnamefont {N.}~\bibnamefont {Noorani}}, \bibinfo {author} {\bibfnamefont {V.}~\bibnamefont {Skavysh}}, \bibinfo {author} {\bibfnamefont {P.}~\bibnamefont {Lamontangne}}, \ and\ \bibinfo {author} {\bibfnamefont {B.~C.}\ \bibnamefont {Sanders}},\ }\href@noop {} {\enquote {\bibinfo {title} {Quantum variational rewinding for time series anomaly detection},}\ } (\bibinfo {year} {2022}),\ \Eprint {http://arxiv.org/abs/arXiv:2210.16438} {arXiv:2210.16438} \BibitemShut {NoStop}%
\bibitem [{\citenamefont {Guo}\ \emph {et~al.}(2023)\citenamefont {Guo}, \citenamefont {Pan}, \citenamefont {Li}, \citenamefont {Gao}, \citenamefont {Qin}, \citenamefont {Yu}, \citenamefont {Zhang},\ and\ \citenamefont {Wen}}]{Guo2023quantumalgo}%
  \BibitemOpen
  \bibfield  {author} {\bibinfo {author} {\bibfnamefont {M.}~\bibnamefont {Guo}}, \bibinfo {author} {\bibfnamefont {S.}~\bibnamefont {Pan}}, \bibinfo {author} {\bibfnamefont {W.}~\bibnamefont {Li}}, \bibinfo {author} {\bibfnamefont {F.}~\bibnamefont {Gao}}, \bibinfo {author} {\bibfnamefont {S.}~\bibnamefont {Qin}}, \bibinfo {author} {\bibfnamefont {X.}~\bibnamefont {Yu}}, \bibinfo {author} {\bibfnamefont {X.}~\bibnamefont {Zhang}}, \ and\ \bibinfo {author} {\bibfnamefont {Q.}~\bibnamefont {Wen}},\ }\href {\doibase https://doi.org/10.1016/j.physa.2023.129018} {\bibfield  {journal} {\bibinfo  {journal} {Physica A: Statistical Mechanics and its Applications}\ }\textbf {\bibinfo {volume} {625}},\ \bibinfo {pages} {129018} (\bibinfo {year} {2023})}\BibitemShut {NoStop}%
\bibitem [{\citenamefont {Llorens}\ \emph {et~al.}(2023)\citenamefont {Llorens}, \citenamefont {Sentís},\ and\ \citenamefont {Muñoz-Tapia}}]{Llorens2023quantummuti}%
  \BibitemOpen
  \bibfield  {author} {\bibinfo {author} {\bibfnamefont {S.}~\bibnamefont {Llorens}}, \bibinfo {author} {\bibfnamefont {G.}~\bibnamefont {Sentís}}, \ and\ \bibinfo {author} {\bibfnamefont {R.}~\bibnamefont {Muñoz-Tapia}},\ }\href@noop {} {\enquote {\bibinfo {title} {Quantum multi-anomaly detection},}\ } (\bibinfo {year} {2023}),\ \Eprint {http://arxiv.org/abs/arXiv:2312.13020} {arXiv:2312.13020} \BibitemShut {NoStop}%
\bibitem [{\citenamefont {Skotiniotis}\ \emph {et~al.}(2018)\citenamefont {Skotiniotis}, \citenamefont {Hotz}, \citenamefont {Calsamiglia},\ and\ \citenamefont {Muñoz-Tapia}}]{Michalis20218Identification}%
  \BibitemOpen
  \bibfield  {author} {\bibinfo {author} {\bibfnamefont {M.}~\bibnamefont {Skotiniotis}}, \bibinfo {author} {\bibfnamefont {R.}~\bibnamefont {Hotz}}, \bibinfo {author} {\bibfnamefont {J.}~\bibnamefont {Calsamiglia}}, \ and\ \bibinfo {author} {\bibfnamefont {R.}~\bibnamefont {Muñoz-Tapia}},\ }\href@noop {} {\enquote {\bibinfo {title} {Identification of malfunctioning quantum devices},}\ } (\bibinfo {year} {2018}),\ \Eprint {http://arxiv.org/abs/arXiv:1808.02729} {arXiv:1808.02729} \BibitemShut {NoStop}%
\bibitem [{\citenamefont {Liu}\ and\ \citenamefont {Rebentrost}(2018)}]{Liu2018QML}%
  \BibitemOpen
  \bibfield  {author} {\bibinfo {author} {\bibfnamefont {N.}~\bibnamefont {Liu}}\ and\ \bibinfo {author} {\bibfnamefont {P.}~\bibnamefont {Rebentrost}},\ }\href {\doibase 10.1103/PhysRevA.97.042315} {\bibfield  {journal} {\bibinfo  {journal} {Phys. Rev. A}\ }\textbf {\bibinfo {volume} {97}},\ \bibinfo {pages} {042315} (\bibinfo {year} {2018})}\BibitemShut {NoStop}%
\bibitem [{\citenamefont {Nielsen}\ and\ \citenamefont {Chuang}(2000)}]{nielsen00}%
  \BibitemOpen
  \bibfield  {author} {\bibinfo {author} {\bibfnamefont {M.~A.}\ \bibnamefont {Nielsen}}\ and\ \bibinfo {author} {\bibfnamefont {I.~L.}\ \bibnamefont {Chuang}},\ }\href@noop {} {\emph {\bibinfo {title} {Quantum Computation and Quantum Information}}}\ (\bibinfo  {publisher} {Cambridge University Press},\ \bibinfo {year} {2000})\BibitemShut {NoStop}%
\bibitem [{\citenamefont {Branderhorst}\ \emph {et~al.}(2009)\citenamefont {Branderhorst}, \citenamefont {Nunn}, \citenamefont {Walmsley},\ and\ \citenamefont {Kosut}}]{Branderhorst2009simplified}%
  \BibitemOpen
  \bibfield  {author} {\bibinfo {author} {\bibfnamefont {M.~P.~A.}\ \bibnamefont {Branderhorst}}, \bibinfo {author} {\bibfnamefont {J.}~\bibnamefont {Nunn}}, \bibinfo {author} {\bibfnamefont {I.~A.}\ \bibnamefont {Walmsley}}, \ and\ \bibinfo {author} {\bibfnamefont {R.~L.}\ \bibnamefont {Kosut}},\ }\href {\doibase 10.1088/1367-2630/11/11/115010} {\bibfield  {journal} {\bibinfo  {journal} {New Journal of Physics}\ }\textbf {\bibinfo {volume} {11}},\ \bibinfo {pages} {115010} (\bibinfo {year} {2009})}\BibitemShut {NoStop}%
\bibitem [{\citenamefont {Shabani}\ \emph {et~al.}(2011)\citenamefont {Shabani}, \citenamefont {Kosut}, \citenamefont {Mohseni}, \citenamefont {Rabitz}, \citenamefont {Broome}, \citenamefont {Almeida}, \citenamefont {Fedrizzi},\ and\ \citenamefont {White}}]{Shabani2011efficient}%
  \BibitemOpen
  \bibfield  {author} {\bibinfo {author} {\bibfnamefont {A.}~\bibnamefont {Shabani}}, \bibinfo {author} {\bibfnamefont {R.~L.}\ \bibnamefont {Kosut}}, \bibinfo {author} {\bibfnamefont {M.}~\bibnamefont {Mohseni}}, \bibinfo {author} {\bibfnamefont {H.}~\bibnamefont {Rabitz}}, \bibinfo {author} {\bibfnamefont {M.~A.}\ \bibnamefont {Broome}}, \bibinfo {author} {\bibfnamefont {M.~P.}\ \bibnamefont {Almeida}}, \bibinfo {author} {\bibfnamefont {A.}~\bibnamefont {Fedrizzi}}, \ and\ \bibinfo {author} {\bibfnamefont {A.~G.}\ \bibnamefont {White}},\ }\href {\doibase 10.1103/PhysRevLett.106.100401} {\bibfield  {journal} {\bibinfo  {journal} {Phys. Rev. Lett.}\ }\textbf {\bibinfo {volume} {106}},\ \bibinfo {pages} {100401} (\bibinfo {year} {2011})}\BibitemShut {NoStop}%
\bibitem [{\citenamefont {DiMario}\ and\ \citenamefont {Becerra}(2021)}]{dimario2021channelnoise}%
  \BibitemOpen
  \bibfield  {author} {\bibinfo {author} {\bibfnamefont {M.~T.}\ \bibnamefont {DiMario}}\ and\ \bibinfo {author} {\bibfnamefont {F.~E.}\ \bibnamefont {Becerra}},\ }\href {\doibase 10.1103/physrevresearch.3.013200} {\bibfield  {journal} {\bibinfo  {journal} {Physical Review Research}\ }\textbf {\bibinfo {volume} {3}} (\bibinfo {year} {2021}),\ 10.1103/physrevresearch.3.013200}\BibitemShut {NoStop}%
\bibitem [{\citenamefont {Sutton}\ and\ \citenamefont {{G. Barto}}(2018)}]{Sutton2018}%
  \BibitemOpen
  \bibfield  {author} {\bibinfo {author} {\bibfnamefont {R.}~\bibnamefont {Sutton}}\ and\ \bibinfo {author} {\bibfnamefont {A.}~\bibnamefont {{G. Barto}}},\ }\href@noop {} {\emph {\bibinfo {title} {{Reinforcement Learning Sutton}}}}\ (\bibinfo  {publisher} {MIT Press},\ \bibinfo {year} {2018})\BibitemShut {NoStop}%
\bibitem [{\citenamefont {Dawid}\ \emph {et~al.}(2022)\citenamefont {Dawid}, \citenamefont {Arnold}, \citenamefont {Requena}, \citenamefont {Gresch}, \citenamefont {Płodzień}, \citenamefont {Donatella}, \citenamefont {Nicoli}, \citenamefont {Stornati}, \citenamefont {Koch}, \citenamefont {Büttner}, \citenamefont {Okuła}, \citenamefont {Muñoz-Gil}, \citenamefont {Vargas-Hernández}, \citenamefont {Cervera-Lierta}, \citenamefont {Carrasquilla}, \citenamefont {Dunjko}, \citenamefont {Gabrié}, \citenamefont {Huembeli}, \citenamefont {van Nieuwenburg}, \citenamefont {Vicentini}, \citenamefont {Wang}, \citenamefont {Wetzel}, \citenamefont {Carleo}, \citenamefont {Greplová}, \citenamefont {Krems}, \citenamefont {Marquardt}, \citenamefont {Tomza}, \citenamefont {Lewenstein},\ and\ \citenamefont {Dauphin}}]{Dawid2022modern}%
  \BibitemOpen
  \bibfield  {author} {\bibinfo {author} {\bibfnamefont {A.}~\bibnamefont {Dawid}}, \bibinfo {author} {\bibfnamefont {J.}~\bibnamefont {Arnold}}, \bibinfo {author} {\bibfnamefont {B.}~\bibnamefont {Requena}}, \bibinfo {author} {\bibfnamefont {A.}~\bibnamefont {Gresch}}, \bibinfo {author} {\bibfnamefont {M.}~\bibnamefont {Płodzień}}, \bibinfo {author} {\bibfnamefont {K.}~\bibnamefont {Donatella}}, \bibinfo {author} {\bibfnamefont {K.~A.}\ \bibnamefont {Nicoli}}, \bibinfo {author} {\bibfnamefont {P.}~\bibnamefont {Stornati}}, \bibinfo {author} {\bibfnamefont {R.}~\bibnamefont {Koch}}, \bibinfo {author} {\bibfnamefont {M.}~\bibnamefont {Büttner}}, \bibinfo {author} {\bibfnamefont {R.}~\bibnamefont {Okuła}}, \bibinfo {author} {\bibfnamefont {G.}~\bibnamefont {Muñoz-Gil}}, \bibinfo {author} {\bibfnamefont {R.~A.}\ \bibnamefont {Vargas-Hernández}}, \bibinfo {author} {\bibfnamefont {A.}~\bibnamefont {Cervera-Lierta}}, \bibinfo {author} {\bibfnamefont {J.}~\bibnamefont {Carrasquilla}}, \bibinfo {author}
  {\bibfnamefont {V.}~\bibnamefont {Dunjko}}, \bibinfo {author} {\bibfnamefont {M.}~\bibnamefont {Gabrié}}, \bibinfo {author} {\bibfnamefont {P.}~\bibnamefont {Huembeli}}, \bibinfo {author} {\bibfnamefont {E.}~\bibnamefont {van Nieuwenburg}}, \bibinfo {author} {\bibfnamefont {F.}~\bibnamefont {Vicentini}}, \bibinfo {author} {\bibfnamefont {L.}~\bibnamefont {Wang}}, \bibinfo {author} {\bibfnamefont {S.~J.}\ \bibnamefont {Wetzel}}, \bibinfo {author} {\bibfnamefont {G.}~\bibnamefont {Carleo}}, \bibinfo {author} {\bibfnamefont {E.}~\bibnamefont {Greplová}}, \bibinfo {author} {\bibfnamefont {R.}~\bibnamefont {Krems}}, \bibinfo {author} {\bibfnamefont {F.}~\bibnamefont {Marquardt}}, \bibinfo {author} {\bibfnamefont {M.}~\bibnamefont {Tomza}}, \bibinfo {author} {\bibfnamefont {M.}~\bibnamefont {Lewenstein}}, \ and\ \bibinfo {author} {\bibfnamefont {A.}~\bibnamefont {Dauphin}},\ }\href@noop {} {\enquote {\bibinfo {title} {Modern applications of machine learning in quantum sciences},}\ } (\bibinfo {year} {2022}),\
  \Eprint {http://arxiv.org/abs/arXiv:2204.04198} {arXiv:2204.04198} \BibitemShut {NoStop}%
\bibitem [{\citenamefont {Borah}\ \emph {et~al.}(2021)\citenamefont {Borah}, \citenamefont {Sarma}, \citenamefont {Kewming}, \citenamefont {Milburn},\ and\ \citenamefont {Twamley}}]{Borah2021measurement}%
  \BibitemOpen
  \bibfield  {author} {\bibinfo {author} {\bibfnamefont {S.}~\bibnamefont {Borah}}, \bibinfo {author} {\bibfnamefont {B.}~\bibnamefont {Sarma}}, \bibinfo {author} {\bibfnamefont {M.}~\bibnamefont {Kewming}}, \bibinfo {author} {\bibfnamefont {G.~J.}\ \bibnamefont {Milburn}}, \ and\ \bibinfo {author} {\bibfnamefont {J.}~\bibnamefont {Twamley}},\ }\href {\doibase 10.1103/PhysRevLett.127.190403} {\bibfield  {journal} {\bibinfo  {journal} {Phys. Rev. Lett.}\ }\textbf {\bibinfo {volume} {127}},\ \bibinfo {pages} {190403} (\bibinfo {year} {2021})}\BibitemShut {NoStop}%
\bibitem [{\citenamefont {Briegel}\ and\ \citenamefont {De~las Cuevas}(2012)}]{Briegel2012projective}%
  \BibitemOpen
  \bibfield  {author} {\bibinfo {author} {\bibfnamefont {H.~J.}\ \bibnamefont {Briegel}}\ and\ \bibinfo {author} {\bibfnamefont {G.}~\bibnamefont {De~las Cuevas}},\ }\href {\doibase 10.1038/srep00400} {\bibfield  {journal} {\bibinfo  {journal} {Scientific Reports}\ }\textbf {\bibinfo {volume} {2}},\ \bibinfo {pages} {400} (\bibinfo {year} {2012})}\BibitemShut {NoStop}%
\bibitem [{\citenamefont {Walln\"ofer}\ \emph {et~al.}(2020)\citenamefont {Walln\"ofer}, \citenamefont {Melnikov}, \citenamefont {D\"ur},\ and\ \citenamefont {Briegel}}]{Wallnofer2020machine}%
  \BibitemOpen
  \bibfield  {author} {\bibinfo {author} {\bibfnamefont {J.}~\bibnamefont {Walln\"ofer}}, \bibinfo {author} {\bibfnamefont {A.~A.}\ \bibnamefont {Melnikov}}, \bibinfo {author} {\bibfnamefont {W.}~\bibnamefont {D\"ur}}, \ and\ \bibinfo {author} {\bibfnamefont {H.~J.}\ \bibnamefont {Briegel}},\ }\href {\doibase 10.1103/PRXQuantum.1.010301} {\bibfield  {journal} {\bibinfo  {journal} {PRX Quantum}\ }\textbf {\bibinfo {volume} {1}},\ \bibinfo {pages} {010301} (\bibinfo {year} {2020})}\BibitemShut {NoStop}%
\bibitem [{\citenamefont {Cui}\ \emph {et~al.}(2022)\citenamefont {Cui}, \citenamefont {Horrocks}, \citenamefont {Hao}, \citenamefont {Guha}, \citenamefont {Peyghambarian}, \citenamefont {Zhuang},\ and\ \citenamefont {Zhang}}]{Cui2022quantum}%
  \BibitemOpen
  \bibfield  {author} {\bibinfo {author} {\bibfnamefont {C.}~\bibnamefont {Cui}}, \bibinfo {author} {\bibfnamefont {W.}~\bibnamefont {Horrocks}}, \bibinfo {author} {\bibfnamefont {S.}~\bibnamefont {Hao}}, \bibinfo {author} {\bibfnamefont {S.}~\bibnamefont {Guha}}, \bibinfo {author} {\bibfnamefont {N.}~\bibnamefont {Peyghambarian}}, \bibinfo {author} {\bibfnamefont {Q.}~\bibnamefont {Zhuang}}, \ and\ \bibinfo {author} {\bibfnamefont {Z.}~\bibnamefont {Zhang}},\ }\href {\doibase 10.1038/s41377-022-01039-5} {\bibfield  {journal} {\bibinfo  {journal} {Light: Science {\&} Applications}\ }\textbf {\bibinfo {volume} {11}},\ \bibinfo {pages} {344} (\bibinfo {year} {2022})}\BibitemShut {NoStop}%
\bibitem [{\citenamefont {Rengaswamy}\ \emph {et~al.}(2021)\citenamefont {Rengaswamy}, \citenamefont {Seshadreesan}, \citenamefont {Guha},\ and\ \citenamefont {Pfister}}]{Rengaswamy2021belief}%
  \BibitemOpen
  \bibfield  {author} {\bibinfo {author} {\bibfnamefont {N.}~\bibnamefont {Rengaswamy}}, \bibinfo {author} {\bibfnamefont {K.~P.}\ \bibnamefont {Seshadreesan}}, \bibinfo {author} {\bibfnamefont {S.}~\bibnamefont {Guha}}, \ and\ \bibinfo {author} {\bibfnamefont {H.~D.}\ \bibnamefont {Pfister}},\ }\href {\doibase 10.1038/s41534-021-00422-1} {\bibfield  {journal} {\bibinfo  {journal} {npj Quantum Information}\ }\textbf {\bibinfo {volume} {7}},\ \bibinfo {pages} {97} (\bibinfo {year} {2021})}\BibitemShut {NoStop}%
\bibitem [{\citenamefont {Piveteau}\ and\ \citenamefont {Renes}(2022)}]{Piveteau2022quantummessage}%
  \BibitemOpen
  \bibfield  {author} {\bibinfo {author} {\bibfnamefont {C.}~\bibnamefont {Piveteau}}\ and\ \bibinfo {author} {\bibfnamefont {J.~M.}\ \bibnamefont {Renes}},\ }\href {\doibase 10.22331/q-2022-08-23-784} {\bibfield  {journal} {\bibinfo  {journal} {{Quantum}}\ }\textbf {\bibinfo {volume} {6}},\ \bibinfo {pages} {784} (\bibinfo {year} {2022})}\BibitemShut {NoStop}%
\bibitem [{\citenamefont {Cortes}\ \emph {et~al.}(2022)\citenamefont {Cortes}, \citenamefont {Lefebvre}, \citenamefont {Lauk}, \citenamefont {Davis}, \citenamefont {Sinclair}, \citenamefont {Gray},\ and\ \citenamefont {Oblak}}]{Cristian2022Sample}%
  \BibitemOpen
  \bibfield  {author} {\bibinfo {author} {\bibfnamefont {C.~L.}\ \bibnamefont {Cortes}}, \bibinfo {author} {\bibfnamefont {P.}~\bibnamefont {Lefebvre}}, \bibinfo {author} {\bibfnamefont {N.}~\bibnamefont {Lauk}}, \bibinfo {author} {\bibfnamefont {M.~J.}\ \bibnamefont {Davis}}, \bibinfo {author} {\bibfnamefont {N.}~\bibnamefont {Sinclair}}, \bibinfo {author} {\bibfnamefont {S.~K.}\ \bibnamefont {Gray}}, \ and\ \bibinfo {author} {\bibfnamefont {D.}~\bibnamefont {Oblak}},\ }\href {journals.aps.org/prapplied/abstract/10.1103/PhysRevApplied.17.034067} {\bibfield  {journal} {\bibinfo  {journal} {Phys. Rev. Applied}\ } (\bibinfo {year} {2022})}\BibitemShut {NoStop}%
\bibitem [{\citenamefont {Sivak}\ \emph {et~al.}(2022)\citenamefont {Sivak}, \citenamefont {Eickbusch}, \citenamefont {Liu}, \citenamefont {Royer}, \citenamefont {Tsioutsios},\ and\ \citenamefont {Devoret}}]{Sivak2022model}%
  \BibitemOpen
  \bibfield  {author} {\bibinfo {author} {\bibfnamefont {V.~V.}\ \bibnamefont {Sivak}}, \bibinfo {author} {\bibfnamefont {A.}~\bibnamefont {Eickbusch}}, \bibinfo {author} {\bibfnamefont {H.}~\bibnamefont {Liu}}, \bibinfo {author} {\bibfnamefont {B.}~\bibnamefont {Royer}}, \bibinfo {author} {\bibfnamefont {I.}~\bibnamefont {Tsioutsios}}, \ and\ \bibinfo {author} {\bibfnamefont {M.~H.}\ \bibnamefont {Devoret}},\ }\href {\doibase 10.1103/PhysRevX.12.011059} {\bibfield  {journal} {\bibinfo  {journal} {Phys. Rev. X}\ }\textbf {\bibinfo {volume} {12}},\ \bibinfo {pages} {011059} (\bibinfo {year} {2022})}\BibitemShut {NoStop}%
\bibitem [{\citenamefont {Niu}\ \emph {et~al.}(2019)\citenamefont {Niu}, \citenamefont {Boixo}, \citenamefont {Smelyanskiy},\ and\ \citenamefont {Neven}}]{Niu2019universal}%
  \BibitemOpen
  \bibfield  {author} {\bibinfo {author} {\bibfnamefont {M.~Y.}\ \bibnamefont {Niu}}, \bibinfo {author} {\bibfnamefont {S.}~\bibnamefont {Boixo}}, \bibinfo {author} {\bibfnamefont {V.~N.}\ \bibnamefont {Smelyanskiy}}, \ and\ \bibinfo {author} {\bibfnamefont {H.}~\bibnamefont {Neven}},\ }\href {\doibase 10.1038/s41534-019-0141-3} {\bibfield  {journal} {\bibinfo  {journal} {npj Quantum Information}\ }\textbf {\bibinfo {volume} {5}},\ \bibinfo {pages} {33} (\bibinfo {year} {2019})}\BibitemShut {NoStop}%
\bibitem [{\citenamefont {F\"osel}\ \emph {et~al.}(2018)\citenamefont {F\"osel}, \citenamefont {Tighineanu}, \citenamefont {Weiss},\ and\ \citenamefont {Marquardt}}]{Fosel2018RL}%
  \BibitemOpen
  \bibfield  {author} {\bibinfo {author} {\bibfnamefont {T.}~\bibnamefont {F\"osel}}, \bibinfo {author} {\bibfnamefont {P.}~\bibnamefont {Tighineanu}}, \bibinfo {author} {\bibfnamefont {T.}~\bibnamefont {Weiss}}, \ and\ \bibinfo {author} {\bibfnamefont {F.}~\bibnamefont {Marquardt}},\ }\href {\doibase 10.1103/PhysRevX.8.031084} {\bibfield  {journal} {\bibinfo  {journal} {Phys. Rev. X}\ }\textbf {\bibinfo {volume} {8}},\ \bibinfo {pages} {031084} (\bibinfo {year} {2018})}\BibitemShut {NoStop}%
\bibitem [{\citenamefont {Altmann}\ \emph {et~al.}(2023)\citenamefont {Altmann}, \citenamefont {Stein}, \citenamefont {Kölle}, \citenamefont {Bärligea}, \citenamefont {Gabor}, \citenamefont {Phan}, \citenamefont {Feld},\ and\ \citenamefont {Linnhoff-Popien}}]{Altmann2023challenges}%
  \BibitemOpen
  \bibfield  {author} {\bibinfo {author} {\bibfnamefont {P.}~\bibnamefont {Altmann}}, \bibinfo {author} {\bibfnamefont {J.}~\bibnamefont {Stein}}, \bibinfo {author} {\bibfnamefont {M.}~\bibnamefont {Kölle}}, \bibinfo {author} {\bibfnamefont {A.}~\bibnamefont {Bärligea}}, \bibinfo {author} {\bibfnamefont {T.}~\bibnamefont {Gabor}}, \bibinfo {author} {\bibfnamefont {T.}~\bibnamefont {Phan}}, \bibinfo {author} {\bibfnamefont {S.}~\bibnamefont {Feld}}, \ and\ \bibinfo {author} {\bibfnamefont {C.}~\bibnamefont {Linnhoff-Popien}},\ }\href@noop {} {\enquote {\bibinfo {title} {Challenges for reinforcement learning in quantum circuit design},}\ } (\bibinfo {year} {2023}),\ \Eprint {http://arxiv.org/abs/arXiv:2312.11337} {arXiv:2312.11337} \BibitemShut {NoStop}%
\bibitem [{\citenamefont {Nägele}\ and\ \citenamefont {Marquardt}(2023)}]{Nagele2023Optimizing}%
  \BibitemOpen
  \bibfield  {author} {\bibinfo {author} {\bibfnamefont {M.}~\bibnamefont {Nägele}}\ and\ \bibinfo {author} {\bibfnamefont {F.}~\bibnamefont {Marquardt}},\ }\href@noop {} {\enquote {\bibinfo {title} {Optimizing zx-diagrams with deep reinforcement learning},}\ } (\bibinfo {year} {2023}),\ \Eprint {http://arxiv.org/abs/arXiv:2311.18588} {arXiv:2311.18588} \BibitemShut {NoStop}%
\bibitem [{\citenamefont {Skolik}\ \emph {et~al.}(2022)\citenamefont {Skolik}, \citenamefont {Cattelan}, \citenamefont {Yarkoni}, \citenamefont {Bäck},\ and\ \citenamefont {Dunjko}}]{Skolik2022Equivariant}%
  \BibitemOpen
  \bibfield  {author} {\bibinfo {author} {\bibfnamefont {A.}~\bibnamefont {Skolik}}, \bibinfo {author} {\bibfnamefont {M.}~\bibnamefont {Cattelan}}, \bibinfo {author} {\bibfnamefont {S.}~\bibnamefont {Yarkoni}}, \bibinfo {author} {\bibfnamefont {T.}~\bibnamefont {Bäck}}, \ and\ \bibinfo {author} {\bibfnamefont {V.}~\bibnamefont {Dunjko}},\ }\href@noop {} {\enquote {\bibinfo {title} {Equivariant quantum circuits for learning on weighted graphs},}\ } (\bibinfo {year} {2022}),\ \Eprint {http://arxiv.org/abs/arXiv:2205.06109} {arXiv:2205.06109} \BibitemShut {NoStop}%
\bibitem [{\citenamefont {Khandelwal}\ and\ \citenamefont {DiAdamo}(2022)}]{Khandelwal2022enhancing}%
  \BibitemOpen
  \bibfield  {author} {\bibinfo {author} {\bibfnamefont {A.}~\bibnamefont {Khandelwal}}\ and\ \bibinfo {author} {\bibfnamefont {S.}~\bibnamefont {DiAdamo}},\ }\href@noop {} {\enquote {\bibinfo {title} {Enhancing protocol privacy with blind calibration of quantum devices},}\ } (\bibinfo {year} {2022}),\ \Eprint {http://arxiv.org/abs/arXiv:2209.05634} {arXiv:2209.05634} \BibitemShut {NoStop}%
\bibitem [{\citenamefont {Zhou}\ \emph {et~al.}(2017)\citenamefont {Zhou}, \citenamefont {Lu}, \citenamefont {Mao}, \citenamefont {yaw Tam},\ and\ \citenamefont {He}}]{Zhou17magnetic}%
  \BibitemOpen
  \bibfield  {author} {\bibinfo {author} {\bibfnamefont {B.}~\bibnamefont {Zhou}}, \bibinfo {author} {\bibfnamefont {C.}~\bibnamefont {Lu}}, \bibinfo {author} {\bibfnamefont {B.-M.}\ \bibnamefont {Mao}}, \bibinfo {author} {\bibfnamefont {H.}~\bibnamefont {yaw Tam}}, \ and\ \bibinfo {author} {\bibfnamefont {S.}~\bibnamefont {He}},\ }\href {\doibase 10.1364/OE.25.008108} {\bibfield  {journal} {\bibinfo  {journal} {Opt. Express}\ }\textbf {\bibinfo {volume} {25}},\ \bibinfo {pages} {8108} (\bibinfo {year} {2017})}\BibitemShut {NoStop}%
\bibitem [{\citenamefont {Cerezo}\ \emph {et~al.}(2021)\citenamefont {Cerezo}, \citenamefont {Arrasmith}, \citenamefont {Babbush}, \citenamefont {Benjamin}, \citenamefont {Endo}, \citenamefont {Fujii}, \citenamefont {McClean}, \citenamefont {Mitarai}, \citenamefont {Yuan}, \citenamefont {Cincio},\ and\ \citenamefont {Coles}}]{Cerezo2021}%
  \BibitemOpen
  \bibfield  {author} {\bibinfo {author} {\bibfnamefont {M.}~\bibnamefont {Cerezo}}, \bibinfo {author} {\bibfnamefont {A.}~\bibnamefont {Arrasmith}}, \bibinfo {author} {\bibfnamefont {R.}~\bibnamefont {Babbush}}, \bibinfo {author} {\bibfnamefont {S.~C.}\ \bibnamefont {Benjamin}}, \bibinfo {author} {\bibfnamefont {S.}~\bibnamefont {Endo}}, \bibinfo {author} {\bibfnamefont {K.}~\bibnamefont {Fujii}}, \bibinfo {author} {\bibfnamefont {J.~R.}\ \bibnamefont {McClean}}, \bibinfo {author} {\bibfnamefont {K.}~\bibnamefont {Mitarai}}, \bibinfo {author} {\bibfnamefont {X.}~\bibnamefont {Yuan}}, \bibinfo {author} {\bibfnamefont {L.}~\bibnamefont {Cincio}}, \ and\ \bibinfo {author} {\bibfnamefont {P.~J.}\ \bibnamefont {Coles}},\ }\href {\doibase 10.1038/s42254-021-00348-9} {\bibfield  {journal} {\bibinfo  {journal} {Nature Reviews Physics}\ }\textbf {\bibinfo {volume} {3}},\ \bibinfo {pages} {625} (\bibinfo {year} {2021})}\BibitemShut {NoStop}%
\bibitem [{\citenamefont {Bharti}\ \emph {et~al.}(2022)\citenamefont {Bharti}, \citenamefont {Cervera-Lierta}, \citenamefont {Kyaw}, \citenamefont {Haug}, \citenamefont {Alperin-Lea}, \citenamefont {Anand}, \citenamefont {Degroote}, \citenamefont {Heimonen}, \citenamefont {Kottmann}, \citenamefont {Menke}, \citenamefont {Mok}, \citenamefont {Sim}, \citenamefont {Kwek},\ and\ \citenamefont {Aspuru-Guzik}}]{NISQreviewAlba}%
  \BibitemOpen
  \bibfield  {author} {\bibinfo {author} {\bibfnamefont {K.}~\bibnamefont {Bharti}}, \bibinfo {author} {\bibfnamefont {A.}~\bibnamefont {Cervera-Lierta}}, \bibinfo {author} {\bibfnamefont {T.~H.}\ \bibnamefont {Kyaw}}, \bibinfo {author} {\bibfnamefont {T.}~\bibnamefont {Haug}}, \bibinfo {author} {\bibfnamefont {S.}~\bibnamefont {Alperin-Lea}}, \bibinfo {author} {\bibfnamefont {A.}~\bibnamefont {Anand}}, \bibinfo {author} {\bibfnamefont {M.}~\bibnamefont {Degroote}}, \bibinfo {author} {\bibfnamefont {H.}~\bibnamefont {Heimonen}}, \bibinfo {author} {\bibfnamefont {J.~S.}\ \bibnamefont {Kottmann}}, \bibinfo {author} {\bibfnamefont {T.}~\bibnamefont {Menke}}, \bibinfo {author} {\bibfnamefont {W.-K.}\ \bibnamefont {Mok}}, \bibinfo {author} {\bibfnamefont {S.}~\bibnamefont {Sim}}, \bibinfo {author} {\bibfnamefont {L.-C.}\ \bibnamefont {Kwek}}, \ and\ \bibinfo {author} {\bibfnamefont {A.}~\bibnamefont {Aspuru-Guzik}},\ }\href {\doibase 10.1103/RevModPhys.94.015004} {\bibfield  {journal} {\bibinfo  {journal} {Rev.
  Mod. Phys.}\ }\textbf {\bibinfo {volume} {94}},\ \bibinfo {pages} {015004} (\bibinfo {year} {2022})}\BibitemShut {NoStop}%
\bibitem [{\citenamefont {Peruzzo}\ \emph {et~al.}(2014)\citenamefont {Peruzzo}, \citenamefont {McClean}, \citenamefont {Shadbolt}, \citenamefont {Yung}, \citenamefont {Zhou}, \citenamefont {Love}, \citenamefont {Aspuru-Guzik},\ and\ \citenamefont {O'Brien}}]{Peruzzo2014quantum}%
  \BibitemOpen
  \bibfield  {author} {\bibinfo {author} {\bibfnamefont {A.}~\bibnamefont {Peruzzo}}, \bibinfo {author} {\bibfnamefont {J.}~\bibnamefont {McClean}}, \bibinfo {author} {\bibfnamefont {P.}~\bibnamefont {Shadbolt}}, \bibinfo {author} {\bibfnamefont {M.-H.}\ \bibnamefont {Yung}}, \bibinfo {author} {\bibfnamefont {X.-Q.}\ \bibnamefont {Zhou}}, \bibinfo {author} {\bibfnamefont {P.~J.}\ \bibnamefont {Love}}, \bibinfo {author} {\bibfnamefont {A.}~\bibnamefont {Aspuru-Guzik}}, \ and\ \bibinfo {author} {\bibfnamefont {J.~L.}\ \bibnamefont {O'Brien}},\ }\href {\doibase 10.1038/ncomms5213} {\bibfield  {journal} {\bibinfo  {journal} {Nature Communications}\ }\textbf {\bibinfo {volume} {5}},\ \bibinfo {pages} {4213} (\bibinfo {year} {2014})}\BibitemShut {NoStop}%
\bibitem [{\citenamefont {Banaszek}\ \emph {et~al.}(2020)\citenamefont {Banaszek}, \citenamefont {Kunz}, \citenamefont {Jachura},\ and\ \citenamefont {Jarzyna}}]{Banaszek2020}%
  \BibitemOpen
  \bibfield  {author} {\bibinfo {author} {\bibfnamefont {K.}~\bibnamefont {Banaszek}}, \bibinfo {author} {\bibfnamefont {L.}~\bibnamefont {Kunz}}, \bibinfo {author} {\bibfnamefont {M.}~\bibnamefont {Jachura}}, \ and\ \bibinfo {author} {\bibfnamefont {M.}~\bibnamefont {Jarzyna}},\ }\href {\doibase 10.1109/JLT.2020.2973890} {\bibfield  {journal} {\bibinfo  {journal} {J. Light. Technol.}\ }\textbf {\bibinfo {volume} {38}},\ \bibinfo {pages} {2741} (\bibinfo {year} {2020})},\ \Eprint {http://arxiv.org/abs/2002.05766} {arXiv:2002.05766} \BibitemShut {NoStop}%
\bibitem [{\citenamefont {Rosati}\ and\ \citenamefont {Giovannetti}(2015)}]{Rosati16b}%
  \BibitemOpen
  \bibfield  {author} {\bibinfo {author} {\bibfnamefont {M.}~\bibnamefont {Rosati}}\ and\ \bibinfo {author} {\bibfnamefont {V.}~\bibnamefont {Giovannetti}},\ }\href {\doibase 10.1063/1.4953690} {\bibfield  {journal} {\bibinfo  {journal} {J. Math. Phys.}\ }\textbf {\bibinfo {volume} {57}},\ \bibinfo {pages} {062204} (\bibinfo {year} {2015})},\ \Eprint {http://arxiv.org/abs/1506.04999} {arXiv:1506.04999} \BibitemShut {NoStop}%
\bibitem [{\citenamefont {Pirandola}\ \emph {et~al.}(2020)\citenamefont {Pirandola}, \citenamefont {Andersen}, \citenamefont {Banchi}, \citenamefont {Berta}, \citenamefont {Bunandar}, \citenamefont {Colbeck}, \citenamefont {Englund}, \citenamefont {Gehring}, \citenamefont {Lupo}, \citenamefont {Ottaviani}, \citenamefont {Pereira}, \citenamefont {Razavi}, \citenamefont {{Shamsul Shaari}}, \citenamefont {Tomamichel}, \citenamefont {Usenko}, \citenamefont {Vallone}, \citenamefont {Villoresi},\ and\ \citenamefont {Wallden}}]{Pirandola2019}%
  \BibitemOpen
  \bibfield  {author} {\bibinfo {author} {\bibfnamefont {S.}~\bibnamefont {Pirandola}}, \bibinfo {author} {\bibfnamefont {U.~L.}\ \bibnamefont {Andersen}}, \bibinfo {author} {\bibfnamefont {L.}~\bibnamefont {Banchi}}, \bibinfo {author} {\bibfnamefont {M.}~\bibnamefont {Berta}}, \bibinfo {author} {\bibfnamefont {D.}~\bibnamefont {Bunandar}}, \bibinfo {author} {\bibfnamefont {R.}~\bibnamefont {Colbeck}}, \bibinfo {author} {\bibfnamefont {D.}~\bibnamefont {Englund}}, \bibinfo {author} {\bibfnamefont {T.}~\bibnamefont {Gehring}}, \bibinfo {author} {\bibfnamefont {C.}~\bibnamefont {Lupo}}, \bibinfo {author} {\bibfnamefont {C.}~\bibnamefont {Ottaviani}}, \bibinfo {author} {\bibfnamefont {J.~L.}\ \bibnamefont {Pereira}}, \bibinfo {author} {\bibfnamefont {M.}~\bibnamefont {Razavi}}, \bibinfo {author} {\bibfnamefont {J.}~\bibnamefont {{Shamsul Shaari}}}, \bibinfo {author} {\bibfnamefont {M.}~\bibnamefont {Tomamichel}}, \bibinfo {author} {\bibfnamefont {V.~C.}\ \bibnamefont {Usenko}}, \bibinfo {author} {\bibfnamefont
  {G.}~\bibnamefont {Vallone}}, \bibinfo {author} {\bibfnamefont {P.}~\bibnamefont {Villoresi}}, \ and\ \bibinfo {author} {\bibfnamefont {P.}~\bibnamefont {Wallden}},\ }\href {\doibase 10.1364/AOP.361502} {\bibfield  {journal} {\bibinfo  {journal} {Adv. Opt. Photonics}\ }\textbf {\bibinfo {volume} {12}},\ \bibinfo {pages} {1012} (\bibinfo {year} {2020})},\ \Eprint {http://arxiv.org/abs/1906.01645} {arXiv:1906.01645} \BibitemShut {NoStop}%
\bibitem [{\citenamefont {Wilde}(2013)}]{wildebook}%
  \BibitemOpen
  \bibfield  {author} {\bibinfo {author} {\bibfnamefont {M.~M.}\ \bibnamefont {Wilde}},\ }\href@noop {} {\emph {\bibinfo {title} {Quantum Information Theory}}}\ (\bibinfo  {publisher} {Cambridge University Press},\ \bibinfo {year} {2013})\BibitemShut {NoStop}%
\bibitem [{\citenamefont {Nasser}\ and\ \citenamefont {Renes}(2017)}]{Nasser2017polar}%
  \BibitemOpen
  \bibfield  {author} {\bibinfo {author} {\bibfnamefont {R.}~\bibnamefont {Nasser}}\ and\ \bibinfo {author} {\bibfnamefont {J.~M.}\ \bibnamefont {Renes}},\ }in\ \href {\doibase 10.1109/ISIT.2017.8006534} {\emph {\bibinfo {booktitle} {2017 IEEE International Symposium on Information Theory (ISIT)}}}\ (\bibinfo {year} {2017})\ pp.\ \bibinfo {pages} {281--285}\BibitemShut {NoStop}%
\bibitem [{\citenamefont {Helstrom}(1976)}]{helstromBOOK}%
  \BibitemOpen
  \bibfield  {author} {\bibinfo {author} {\bibfnamefont {C.~W.}\ \bibnamefont {Helstrom}},\ }\href@noop {} {\emph {\bibinfo {title} {{Quantum Detection and Estimation Theory}}}},\ Vol.~\bibinfo {volume} {84}\ (\bibinfo  {publisher} {Academic press},\ \bibinfo {address} {New York},\ \bibinfo {year} {1976})\ p.\ \bibinfo {pages} {309}\BibitemShut {NoStop}%
\bibitem [{\citenamefont {Assalini}\ \emph {et~al.}(2011)\citenamefont {Assalini}, \citenamefont {Dalla~Pozza},\ and\ \citenamefont {Pierobon}}]{Assalini2011revisting}%
  \BibitemOpen
  \bibfield  {author} {\bibinfo {author} {\bibfnamefont {A.}~\bibnamefont {Assalini}}, \bibinfo {author} {\bibfnamefont {N.}~\bibnamefont {Dalla~Pozza}}, \ and\ \bibinfo {author} {\bibfnamefont {G.}~\bibnamefont {Pierobon}},\ }\href {\doibase 10.1103/PhysRevA.84.022342} {\bibfield  {journal} {\bibinfo  {journal} {Phys. Rev. A}\ }\textbf {\bibinfo {volume} {84}},\ \bibinfo {pages} {022342} (\bibinfo {year} {2011})}\BibitemShut {NoStop}%
\bibitem [{\citenamefont {Zoratti}\ \emph {et~al.}(2021)\citenamefont {Zoratti}, \citenamefont {Dalla~Pozza}, \citenamefont {Fanizza},\ and\ \citenamefont {Giovannetti}}]{Zoratti2021agnostic}%
  \BibitemOpen
  \bibfield  {author} {\bibinfo {author} {\bibfnamefont {F.}~\bibnamefont {Zoratti}}, \bibinfo {author} {\bibfnamefont {N.}~\bibnamefont {Dalla~Pozza}}, \bibinfo {author} {\bibfnamefont {M.}~\bibnamefont {Fanizza}}, \ and\ \bibinfo {author} {\bibfnamefont {V.}~\bibnamefont {Giovannetti}},\ }\href {\doibase 10.1103/PhysRevA.104.042606} {\bibfield  {journal} {\bibinfo  {journal} {Phys. Rev. A}\ }\textbf {\bibinfo {volume} {104}},\ \bibinfo {pages} {042606} (\bibinfo {year} {2021})}\BibitemShut {NoStop}%
\bibitem [{\citenamefont {Takeoka}\ \emph {et~al.}(2005)\citenamefont {Takeoka}, \citenamefont {Sasaki}, \citenamefont {van Loock},\ and\ \citenamefont {L\"utkenhaus}}]{Takeoka2005Implementation}%
  \BibitemOpen
  \bibfield  {author} {\bibinfo {author} {\bibfnamefont {M.}~\bibnamefont {Takeoka}}, \bibinfo {author} {\bibfnamefont {M.}~\bibnamefont {Sasaki}}, \bibinfo {author} {\bibfnamefont {P.}~\bibnamefont {van Loock}}, \ and\ \bibinfo {author} {\bibfnamefont {N.}~\bibnamefont {L\"utkenhaus}},\ }\href {\doibase 10.1103/PhysRevA.71.022318} {\bibfield  {journal} {\bibinfo  {journal} {Phys. Rev. A}\ }\textbf {\bibinfo {volume} {71}},\ \bibinfo {pages} {022318} (\bibinfo {year} {2005})}\BibitemShut {NoStop}%
\bibitem [{\citenamefont {Cook}\ \emph {et~al.}(2007)\citenamefont {Cook}, \citenamefont {Martin}, \citenamefont {Geremia}, \citenamefont {Chase},\ and\ \citenamefont {Geremia}}]{Cook2007}%
  \BibitemOpen
  \bibfield  {author} {\bibinfo {author} {\bibfnamefont {R.~L.}\ \bibnamefont {Cook}}, \bibinfo {author} {\bibfnamefont {P.~J.}\ \bibnamefont {Martin}}, \bibinfo {author} {\bibfnamefont {J.~M.}\ \bibnamefont {Geremia}}, \bibinfo {author} {\bibfnamefont {B.~A.}\ \bibnamefont {Chase}}, \ and\ \bibinfo {author} {\bibfnamefont {J.~M.}\ \bibnamefont {Geremia}},\ }\href {\doibase 10.1038/nature05655} {\bibfield  {journal} {\bibinfo  {journal} {Nature}\ }\textbf {\bibinfo {volume} {446}},\ \bibinfo {pages} {774} (\bibinfo {year} {2007})}\BibitemShut {NoStop}%
\bibitem [{\citenamefont {Sych}\ and\ \citenamefont {Leuchs}(2016)}]{Sych2016practical}%
  \BibitemOpen
  \bibfield  {author} {\bibinfo {author} {\bibfnamefont {D.}~\bibnamefont {Sych}}\ and\ \bibinfo {author} {\bibfnamefont {G.}~\bibnamefont {Leuchs}},\ }\href {\doibase 10.1103/PhysRevLett.117.200501} {\bibfield  {journal} {\bibinfo  {journal} {Phys. Rev. Lett.}\ }\textbf {\bibinfo {volume} {117}},\ \bibinfo {pages} {200501} (\bibinfo {year} {2016})}\BibitemShut {NoStop}%
\bibitem [{\citenamefont {Becerra}\ \emph {et~al.}(2011)\citenamefont {Becerra}, \citenamefont {Fan}, \citenamefont {Baumgartner}, \citenamefont {Polyakov}, \citenamefont {Goldhar}, \citenamefont {Kosloski},\ and\ \citenamefont {Migdall}}]{becerra2011mary}%
  \BibitemOpen
  \bibfield  {author} {\bibinfo {author} {\bibfnamefont {F.~E.}\ \bibnamefont {Becerra}}, \bibinfo {author} {\bibfnamefont {J.}~\bibnamefont {Fan}}, \bibinfo {author} {\bibfnamefont {G.}~\bibnamefont {Baumgartner}}, \bibinfo {author} {\bibfnamefont {S.~V.}\ \bibnamefont {Polyakov}}, \bibinfo {author} {\bibfnamefont {J.}~\bibnamefont {Goldhar}}, \bibinfo {author} {\bibfnamefont {J.~T.}\ \bibnamefont {Kosloski}}, \ and\ \bibinfo {author} {\bibfnamefont {A.}~\bibnamefont {Migdall}},\ }\href {\doibase 10.1103/PhysRevA.84.062324} {\bibfield  {journal} {\bibinfo  {journal} {Phys. Rev. A}\ }\textbf {\bibinfo {volume} {84}},\ \bibinfo {pages} {062324} (\bibinfo {year} {2011})}\BibitemShut {NoStop}%
\bibitem [{\citenamefont {Dolinar}(1973)}]{Dolinar1973}%
  \BibitemOpen
  \bibfield  {author} {\bibinfo {author} {\bibfnamefont {S.~J.}\ \bibnamefont {Dolinar}},\ }\href {https://dspace.mit.edu/handle/1721.1/56414} {\bibfield  {journal} {\bibinfo  {journal} {Q. Prog. Rep. (Research Lab. Electron.}\ }\textbf {\bibinfo {volume} {111}},\ \bibinfo {pages} {115} (\bibinfo {year} {1973})}\BibitemShut {NoStop}%
\bibitem [{\citenamefont {Usenko}\ \emph {et~al.}(2012)\citenamefont {Usenko}, \citenamefont {Heim}, \citenamefont {Peuntinger}, \citenamefont {Wittmann}, \citenamefont {Marquardt}, \citenamefont {Leuchs},\ and\ \citenamefont {Filip}}]{Usenko2012a}%
  \BibitemOpen
  \bibfield  {author} {\bibinfo {author} {\bibfnamefont {V.~C.}\ \bibnamefont {Usenko}}, \bibinfo {author} {\bibfnamefont {B.}~\bibnamefont {Heim}}, \bibinfo {author} {\bibfnamefont {C.}~\bibnamefont {Peuntinger}}, \bibinfo {author} {\bibfnamefont {C.}~\bibnamefont {Wittmann}}, \bibinfo {author} {\bibfnamefont {C.}~\bibnamefont {Marquardt}}, \bibinfo {author} {\bibfnamefont {G.}~\bibnamefont {Leuchs}}, \ and\ \bibinfo {author} {\bibfnamefont {R.}~\bibnamefont {Filip}},\ }\href {\doibase 10.1088/1367-2630/14/9/093048} {\bibfield  {journal} {\bibinfo  {journal} {New J. Phys.}\ }\textbf {\bibinfo {volume} {14}},\ \bibinfo {pages} {093048} (\bibinfo {year} {2012})}\BibitemShut {NoStop}%
\bibitem [{\citenamefont {DiMario}\ and\ \citenamefont {Becerra}(2022)}]{DiMario2022}%
  \BibitemOpen
  \bibfield  {author} {\bibinfo {author} {\bibfnamefont {M.~T.}\ \bibnamefont {DiMario}}\ and\ \bibinfo {author} {\bibfnamefont {F.~E.}\ \bibnamefont {Becerra}},\ }\href {\doibase 10.1038/s41534-022-00595-3} {\bibfield  {journal} {\bibinfo  {journal} {npj Quantum Information}\ }\textbf {\bibinfo {volume} {8}},\ \bibinfo {pages} {84} (\bibinfo {year} {2022})}\BibitemShut {NoStop}%
\bibitem [{\citenamefont {Kennedy}(1973)}]{Kennedy1973a}%
  \BibitemOpen
  \bibfield  {author} {\bibinfo {author} {\bibfnamefont {R.~S.}\ \bibnamefont {Kennedy}},\ }\href {https://dspace.mit.edu/handle/1721.1/56346} {\bibfield  {journal} {\bibinfo  {journal} {MIT Res. Lab. Electron. Q. Prog. Rep.}\ }\textbf {\bibinfo {volume} {108}},\ \bibinfo {pages} {219} (\bibinfo {year} {1973})}\BibitemShut {NoStop}%
\bibitem [{\citenamefont {Ferraro}\ \emph {et~al.}(2005)\citenamefont {Ferraro}, \citenamefont {Olivares},\ and\ \citenamefont {Paris}}]{RevGauss}%
  \BibitemOpen
  \bibfield  {author} {\bibinfo {author} {\bibfnamefont {A.}~\bibnamefont {Ferraro}}, \bibinfo {author} {\bibfnamefont {S.}~\bibnamefont {Olivares}}, \ and\ \bibinfo {author} {\bibfnamefont {M.~G.~A.}\ \bibnamefont {Paris}},\ }\href@noop {} {\emph {\bibinfo {title} {{Gaussian states in continuous variable quantum information}}}}\ (\bibinfo  {publisher} {Bibliopolis},\ \bibinfo {address} {Napoli},\ \bibinfo {year} {2005})\BibitemShut {NoStop}%
\bibitem [{\citenamefont {PAGE}(1955)}]{Page1955test}%
  \BibitemOpen
  \bibfield  {author} {\bibinfo {author} {\bibfnamefont {E.~S.}\ \bibnamefont {PAGE}},\ }\href {\doibase 10.1093/biomet/42.3-4.523} {\bibfield  {journal} {\bibinfo  {journal} {Biometrika}\ }\textbf {\bibinfo {volume} {42}},\ \bibinfo {pages} {523} (\bibinfo {year} {1955})},\ \Eprint {http://arxiv.org/abs/https://academic.oup.com/biomet/article-pdf/42/3-4/523/838813/42-3-4-523.pdf} {https://academic.oup.com/biomet/article-pdf/42/3-4/523/838813/42-3-4-523.pdf} \BibitemShut {NoStop}%
\bibitem [{\citenamefont {Brodsky}\ and\ \citenamefont {Darkhovsky}(2010)}]{brodsky2010non}%
  \BibitemOpen
  \bibfield  {author} {\bibinfo {author} {\bibfnamefont {E.}~\bibnamefont {Brodsky}}\ and\ \bibinfo {author} {\bibfnamefont {B.}~\bibnamefont {Darkhovsky}},\ }\href {https://books.google.es/books?id=Ar56cgAACAAJ} {\emph {\bibinfo {title} {Non-Parametric Statistical Diagnosis: Problems and Methods}}},\ Mathematics and Its Applications\ (\bibinfo  {publisher} {Springer Netherlands},\ \bibinfo {year} {2010})\BibitemShut {NoStop}%
\bibitem [{\citenamefont {Basseville}\ and\ \citenamefont {Nikiforov}(1993)}]{Basseville1993detection}%
  \BibitemOpen
  \bibfield  {author} {\bibinfo {author} {\bibfnamefont {M.}~\bibnamefont {Basseville}}\ and\ \bibinfo {author} {\bibfnamefont {I.}~\bibnamefont {Nikiforov}},\ }\href@noop {} {\emph {\bibinfo {title} {Detection of Abrupt Change Theory and Application}}},\ Vol.~\bibinfo {volume} {15}\ (\bibinfo  {publisher} {Prentice Hall},\ \bibinfo {year} {1993})\BibitemShut {NoStop}%
\bibitem [{\citenamefont {Abel}\ \emph {et~al.}(2023)\citenamefont {Abel}, \citenamefont {Barreto}, \citenamefont {Roy}, \citenamefont {Precup}, \citenamefont {van Hasselt},\ and\ \citenamefont {Singh}}]{abel2023definition}%
  \BibitemOpen
  \bibfield  {author} {\bibinfo {author} {\bibfnamefont {D.}~\bibnamefont {Abel}}, \bibinfo {author} {\bibfnamefont {A.}~\bibnamefont {Barreto}}, \bibinfo {author} {\bibfnamefont {B.~V.}\ \bibnamefont {Roy}}, \bibinfo {author} {\bibfnamefont {D.}~\bibnamefont {Precup}}, \bibinfo {author} {\bibfnamefont {H.}~\bibnamefont {van Hasselt}}, \ and\ \bibinfo {author} {\bibfnamefont {S.}~\bibnamefont {Singh}},\ }\href@noop {} {\enquote {\bibinfo {title} {A definition of continual reinforcement learning},}\ } (\bibinfo {year} {2023}),\ \Eprint {http://arxiv.org/abs/arXiv:2307.11046} {arXiv:2307.11046} \BibitemShut {NoStop}%
\bibitem [{\citenamefont {Khetarpal}\ \emph {et~al.}(2020)\citenamefont {Khetarpal}, \citenamefont {Riemer}, \citenamefont {Rish},\ and\ \citenamefont {Precup}}]{khetarpal2020towards}%
  \BibitemOpen
  \bibfield  {author} {\bibinfo {author} {\bibfnamefont {K.}~\bibnamefont {Khetarpal}}, \bibinfo {author} {\bibfnamefont {M.}~\bibnamefont {Riemer}}, \bibinfo {author} {\bibfnamefont {I.}~\bibnamefont {Rish}}, \ and\ \bibinfo {author} {\bibfnamefont {D.}~\bibnamefont {Precup}},\ }\href@noop {} {\enquote {\bibinfo {title} {Towards continual reinforcement learning: A review and perspectives},}\ } (\bibinfo {year} {2020}),\ \Eprint {http://arxiv.org/abs/arXiv:2012.13490} {arXiv:2012.13490} \BibitemShut {NoStop}%
\bibitem [{\citenamefont {Tobin}\ \emph {et~al.}(2017)\citenamefont {Tobin}, \citenamefont {Fong}, \citenamefont {Ray}, \citenamefont {Schneider}, \citenamefont {Zaremba},\ and\ \citenamefont {Abbeel}}]{tobin2017domain}%
  \BibitemOpen
  \bibfield  {author} {\bibinfo {author} {\bibfnamefont {J.}~\bibnamefont {Tobin}}, \bibinfo {author} {\bibfnamefont {R.}~\bibnamefont {Fong}}, \bibinfo {author} {\bibfnamefont {A.}~\bibnamefont {Ray}}, \bibinfo {author} {\bibfnamefont {J.}~\bibnamefont {Schneider}}, \bibinfo {author} {\bibfnamefont {W.}~\bibnamefont {Zaremba}}, \ and\ \bibinfo {author} {\bibfnamefont {P.}~\bibnamefont {Abbeel}},\ }in\ \href {\doibase 10.1109/IROS.2017.8202133} {\emph {\bibinfo {booktitle} {2017 IEEE/RSJ International Conference on Intelligent Robots and Systems (IROS)}}}\ (\bibinfo {year} {2017})\ pp.\ \bibinfo {pages} {23--30}\BibitemShut {NoStop}%
\bibitem [{\citenamefont {Xing}\ \emph {et~al.}(2021)\citenamefont {Xing}, \citenamefont {Nagata}, \citenamefont {Chen}, \citenamefont {Zou}, \citenamefont {Neftci},\ and\ \citenamefont {Krichmar}}]{xing2021domain}%
  \BibitemOpen
  \bibfield  {author} {\bibinfo {author} {\bibfnamefont {J.}~\bibnamefont {Xing}}, \bibinfo {author} {\bibfnamefont {T.}~\bibnamefont {Nagata}}, \bibinfo {author} {\bibfnamefont {K.}~\bibnamefont {Chen}}, \bibinfo {author} {\bibfnamefont {X.}~\bibnamefont {Zou}}, \bibinfo {author} {\bibfnamefont {E.}~\bibnamefont {Neftci}}, \ and\ \bibinfo {author} {\bibfnamefont {J.~L.}\ \bibnamefont {Krichmar}},\ }\href@noop {} {\enquote {\bibinfo {title} {Domain adaptation in reinforcement learning via latent unified state representation},}\ } (\bibinfo {year} {2021}),\ \Eprint {http://arxiv.org/abs/arXiv:2102.05714} {arXiv:2102.05714} \BibitemShut {NoStop}%
\bibitem [{\citenamefont {Radeva}\ \emph {et~al.}(2012)\citenamefont {Radeva}, \citenamefont {Drozdzal}, \citenamefont {Segui}, \citenamefont {Igual}, \citenamefont {Malagelada}, \citenamefont {Azpiroz},\ and\ \citenamefont {Vitria}}]{Dadeva2012active}%
  \BibitemOpen
  \bibfield  {author} {\bibinfo {author} {\bibfnamefont {P.}~\bibnamefont {Radeva}}, \bibinfo {author} {\bibfnamefont {M.}~\bibnamefont {Drozdzal}}, \bibinfo {author} {\bibfnamefont {S.}~\bibnamefont {Segui}}, \bibinfo {author} {\bibfnamefont {L.}~\bibnamefont {Igual}}, \bibinfo {author} {\bibfnamefont {C.}~\bibnamefont {Malagelada}}, \bibinfo {author} {\bibfnamefont {F.}~\bibnamefont {Azpiroz}}, \ and\ \bibinfo {author} {\bibfnamefont {J.}~\bibnamefont {Vitria}},\ }in\ \href {\doibase 10.1109/HPCSim.2012.6266908} {\emph {\bibinfo {booktitle} {2012 International Conference on High Performance Computing \& Simulation (HPCS)}}}\ (\bibinfo {year} {2012})\ pp.\ \bibinfo {pages} {174--181}\BibitemShut {NoStop}%
\bibitem [{\citenamefont {Baum}\ \emph {et~al.}(2021)\citenamefont {Baum}, \citenamefont {Amico}, \citenamefont {Howell}, \citenamefont {Hush}, \citenamefont {Liuzzi}, \citenamefont {Mundada}, \citenamefont {Merkh}, \citenamefont {Carvalho},\ and\ \citenamefont {Biercuk}}]{Baum2021Experimental}%
  \BibitemOpen
  \bibfield  {author} {\bibinfo {author} {\bibfnamefont {Y.}~\bibnamefont {Baum}}, \bibinfo {author} {\bibfnamefont {M.}~\bibnamefont {Amico}}, \bibinfo {author} {\bibfnamefont {S.}~\bibnamefont {Howell}}, \bibinfo {author} {\bibfnamefont {M.}~\bibnamefont {Hush}}, \bibinfo {author} {\bibfnamefont {M.}~\bibnamefont {Liuzzi}}, \bibinfo {author} {\bibfnamefont {P.}~\bibnamefont {Mundada}}, \bibinfo {author} {\bibfnamefont {T.}~\bibnamefont {Merkh}}, \bibinfo {author} {\bibfnamefont {A.~R.}\ \bibnamefont {Carvalho}}, \ and\ \bibinfo {author} {\bibfnamefont {M.~J.}\ \bibnamefont {Biercuk}},\ }\href {\doibase 10.1103/PRXQuantum.2.040324} {\bibfield  {journal} {\bibinfo  {journal} {PRX Quantum}\ }\textbf {\bibinfo {volume} {2}},\ \bibinfo {pages} {040324} (\bibinfo {year} {2021})}\BibitemShut {NoStop}%
\bibitem [{\citenamefont {Liao}\ \emph {et~al.}(2021)\citenamefont {Liao}, \citenamefont {Hsieh},\ and\ \citenamefont {Ferrie}}]{Liao2021quantum}%
  \BibitemOpen
  \bibfield  {author} {\bibinfo {author} {\bibfnamefont {Y.}~\bibnamefont {Liao}}, \bibinfo {author} {\bibfnamefont {M.-H.}\ \bibnamefont {Hsieh}}, \ and\ \bibinfo {author} {\bibfnamefont {C.}~\bibnamefont {Ferrie}},\ }\href@noop {} {\enquote {\bibinfo {title} {Quantum optimization for training quantum neural networks},}\ } (\bibinfo {year} {2021}),\ \Eprint {http://arxiv.org/abs/arXiv:2103.17047} {arXiv:2103.17047} \BibitemShut {NoStop}%
\bibitem [{\citenamefont {et. al.}(2023)}]{abbas2023quantumopt}%
  \BibitemOpen
  \bibfield  {author} {\bibinfo {author} {\bibfnamefont {A.~A.}\ \bibnamefont {et. al.}},\ }\href@noop {} {\enquote {\bibinfo {title} {Quantum optimization: Potential, challenges, and the path forward},}\ } (\bibinfo {year} {2023}),\ \Eprint {http://arxiv.org/abs/arXiv:2312.02279} {arXiv:2312.02279} \BibitemShut {NoStop}%
\bibitem [{\citenamefont {BELLMAN}\ and\ \citenamefont {Dreyfus}(2010)}]{bellmannDP}%
  \BibitemOpen
  \bibfield  {author} {\bibinfo {author} {\bibfnamefont {R.}~\bibnamefont {BELLMAN}}\ and\ \bibinfo {author} {\bibfnamefont {S.}~\bibnamefont {Dreyfus}},\ }\href {http://www.jstor.org/stable/j.ctv1nxcw0f} {\emph {\bibinfo {title} {Dynamic Programming}}},\ Vol.~\bibinfo {volume} {33}\ (\bibinfo  {publisher} {Princeton University Press},\ \bibinfo {year} {2010})\BibitemShut {NoStop}%
\bibitem [{\citenamefont {Szepesv{\'a}ri}(2010)}]{algsrl}%
  \BibitemOpen
  \bibfield  {author} {\bibinfo {author} {\bibfnamefont {C.}~\bibnamefont {Szepesv{\'a}ri}},\ }\href {http://dx.doi.org/10.2200/S00268ED1V01Y201005AIM009} {\emph {\bibinfo {title} {Algorithms for Reinforcement Learning}}},\ Synthesis lectures on artificial intelligence and machine learning\ (\bibinfo  {publisher} {Morgan \& Claypool},\ \bibinfo {year} {2010})\BibitemShut {NoStop}%
\bibitem [{\citenamefont {Watkins}(1989)}]{Watkins1989}%
  \BibitemOpen
  \bibfield  {author} {\bibinfo {author} {\bibfnamefont {C.}~\bibnamefont {Watkins}},\ }\emph {\bibinfo {title} {{“Learning from delayed rewards”. PhD thesis}}},\ \href {http://www.cs.rhul.ac.uk/{~}chrisw/thesis.html} {Ph.D. thesis},\ \bibinfo  {school} {Cambridge} (\bibinfo {year} {1989})\BibitemShut {NoStop}%
\bibitem [{\citenamefont {Lattimore}\ and\ \citenamefont {Szepesvári}(2020)}]{Lattimore_Szepesvári_2020}%
  \BibitemOpen
  \bibfield  {author} {\bibinfo {author} {\bibfnamefont {T.}~\bibnamefont {Lattimore}}\ and\ \bibinfo {author} {\bibfnamefont {C.}~\bibnamefont {Szepesvári}},\ }\href@noop {} {\emph {\bibinfo {title} {Bandit Algorithms}}}\ (\bibinfo  {publisher} {Cambridge University Press},\ \bibinfo {year} {2020})\BibitemShut {NoStop}%
\bibitem [{\citenamefont {Crosta}\ \emph {et~al.}(2024)\citenamefont {Crosta}, \citenamefont {Matera},\ and\ \citenamefont {Bilkis}}]{Crosta2024Repo}%
  \BibitemOpen
  \bibfield  {author} {\bibinfo {author} {\bibfnamefont {T.}~\bibnamefont {Crosta}}, \bibinfo {author} {\bibfnamefont {M.}~\bibnamefont {Matera}}, \ and\ \bibinfo {author} {\bibfnamefont {M.}~\bibnamefont {Bilkis}},\ }\href {https://github.com/dmtomas/qrec} {\emph {\bibinfo {title} {Repository}}}\ (\bibinfo  {publisher} {GitHub},\ \bibinfo {year} {2024})\BibitemShut {NoStop}%
\end{thebibliography}%

\appendix

\section{Additional details in the RL implementation}
\label{chap:append}
In the following we briefly discuss the Reinforcement Learning (RL) framework and provide further details on the illustrative example considered in Sec.~\ref{sec:results}.

\textit{Reinforcement Learning} is based on the sequential interaction between an agent and the environment during several episodes~\cite{Sutton2018}. Each episode $E$ consists on steps $t=1,...,T$ (where $T$ is potentially of stochastic nature). At step $t$ the agent observes a \textit{state} $s_t$, and follows a policy $\pi(a_t|s_t)\equiv\pi$ in order to choose an \textit{action} $a_t$. As a consequence, the agent receives a \textit{reward} $r_{t+1}$ and transitions to the next state $s_{t+1}$. The goal of the agent is to maximize the reward acquired during episodes, which is accomplished by performing the \textit{optimal policy}. To do this, the agent has to exploit valuable actions but also explore possibly advantageous configurations, leading to an exploration-exploitation trade-off. Since intermediate rewards might be enjoyed during an episode, the \textit{return} is defined as $G_{t}=\sum_{k=0}^{T-t}\gamma^k r_{t+k+1}$, where $\gamma$ is a weighting factor. Here, we note that the return is a quantity that depends on the sequence of state and actions visited during each episode. Thus, the average value of the return, $Q_\pi(s,a)=\E{\pi}{G_t | s_t=s, a_t=a}$ serves as a measure of how valuable it is to take action $a$ starting from the state $s$ and following policy $\pi$ thereafter. Here, the optimal policy $\pi^*$ can be obtained by finding the maximum Q-value for each given state, a problem that is reflected by the so-called \textit{optimal Bellman equation}~\cite{Sutton2018,bellmannDP}. 

The Q-learning algorithm exploits the structure of Bellman equations by linking them to contractive operations and shifting the policy towards the fixed point associated to the optimal Bellman operator~\cite{algsrl,Watkins1989}. In order to find the optimal Q-values $Q^{*}(s,a) = Q_{\pi^*}(s,a)$, Q-learning updates the $Q$-estimate as
{\small\be\begin{aligned} \label{eq:QLUPDATERULE}
\hat{Q}(s_{t}, a_t)  \leftarrow (1-\lambda_E)\hat{Q}(s_t, a_t)
&+ \lambda_E \left(r_{t+1}  + \gamma \max_{a'}\hat{Q}(s_{t+1}, a')\right),
\end{aligned}\ee}%
with $\lambda_E$ an episode-dependent learning-rate; in Alg.~\ref{alg:ql} we sketch the Q-learning pseudo-code. Here, the agent explores the state-action space by committing to an $\epsilon$-greedy policy $\pi_\epsilon$, where a random action is chosen with probability $\epsilon$, and the action which maximizes the current state-action value estimate $\hat{Q}(s,a)$ is chosen otherwise. Note that \textit{(i)} such \textit{greedy} action might potentially be sub-optimal option, and \textit{(ii)} a schedule for $\epsilon$ is set set in practice, in order to balance between exploration and exploitation~\cite{Lattimore_Szepesvári_2020,Sutton2018,bilkis2020real}.

\begin{table}[t!]
\centering
\begin{tabular}{||p{1.8cm}| p{3.2cm} |p{1.5cm} |p{1.5cm}||}
 \hline
 Parameters & Meaning & Proposed method & Q-learning \\ [0.5ex] 
 \hline
check jump threshold & How much repeated selection of the maximum is considered conversion. & 3000 & 3000 \\ 
\hline
$\delta$ & How much the change in $\Wd$ has to be in order to recalibrate. & 0.1 & 0.1 \\ 
 \hline
 $\epsilon_0$ & Minimum exploration of the agent. & 0.05 & 0.1 \\
 \hline
 $\Delta_\epsilon$ & Rate of change for $\epsilon$. & 0.9 & 0.9999 \\
 \hline
 $\Delta$ & How much do we want to deviate from a uniform distribution. & 50 & 0 \\
 \hline
 $\Delta_l$ & Step from which the learning rate starts  & 150 & 1 \\ [1ex] 
 \hline
\end{tabular}
\caption{Hyperparameters used in the numerical examples with a description of its interpretations.}
\label{table:hyperparams}
\end{table}

\begin{algorithm}[ht]\label{alg:ql}
  \DontPrintSemicolon
  \SetAlgoNoEnd
  \For{ \texttt{episode} E = 1, ... }{
  $\; \; \;$ \texttt{initialize} $ s_0  \; \; \;$ \;
  $\; \; \;$ \For{\texttt{step }$t$ \texttt{in episode} $E$}{$\; \; \; \; \; \;$\texttt{choose }$a_t \sim \pi_\epsilon$\; $\; \; \; \; \; \;$\texttt{get }$r_{t},s_{t+1}$ \; $\; \; \; \; \; \;  $\texttt{update }$\hat{Q}(s_t, a_t)$ using Eq.~\ref{eq:QLUPDATERULE}. 
  }
}
\caption{Q-learning pseudocode.}
\end{algorithm}

\vspace{0.3cm}
We now turn to provide additional details on the numerical implementation for the Kennedy receiver considered in Sec.~\ref{sec:results}. Our code is open-sourced and can be found in Ref.~\cite{Crosta2024Repo}.

\textit{De-calibration witness}. In order to detect changes in the environment during an off-calibration stage, the average output of the detector is computed across experiments $E=1,...,N_{\text{eff}}$, where we set $N_{\text{eff}}=1000$. In other words, ${\Wd}^{(E)} = \frac{1}{N_{\text{eff}}} \sum_{i=0}^{N_{\text{eff}}} n_{t-i}$. At each experiment, the difference between the current average and the previous one is computed $|{\Wd}^{E} - {\Wd}^{E-1}|$. Here, if this difference is bigger than the (hyper)parameter $\delta$ --- and assuming the device is being deployed --- the re-calibration process is restarted.

\textit{Effective model}. When the (re-)calibration is initiated, we consider an effective model given by the success probability of Kennedy receiver, which can be computed if  estimating signal's intensity. Such success probability can be linked to the optimal Q-values for an ideal environment $\mathcal{E}_0$ in which the device functions correctly~\cite{bilkis2020real}. Thus, the effective model is set by estimating $|\alpha|^2$, which serves as an internal model parameter. To this end, the displacement value in the Kennedy receiver is set to zero during $N_{\text{eff}}=1000$ experiments, and the intensity $|\alpha|^2$ is estimated as per $|\alpha|^2 = -\ln(p(n=0\big|\alpha)) \pm \frac{1}{N}$. Consequently, the Q-learning agent fine-tunes the device configuration, which is potentially deployed under an environment whose score function differs from the noiseless effective-model here considered. Importantly, the Q-values are initialized to $Q_0^{\alpha}\left(\theta\right)$ and $Q_1^{\alpha}(\theta, \hat{k})$ as per $
Q_0^{\alpha}(\theta) = \sum_{n=\{0, 1\}} \max_{\hat{k}={0,1}}Q_1^{\alpha}(\theta, \hat{k})$ and $
    Q_1^{\alpha}(\theta, \hat{k}) = \frac{1}{2} e^{-|(-1)^{\hat{k}} \alpha + \theta|^2}+ \frac{1}{2} \left(1- e^{-|(-1)^{\hat{k} + 1}\alpha + \theta|^2}\right)$
\vspace{0.1cm}

\begin{figure}[t!]
\centering
\includegraphics[scale=0.27]{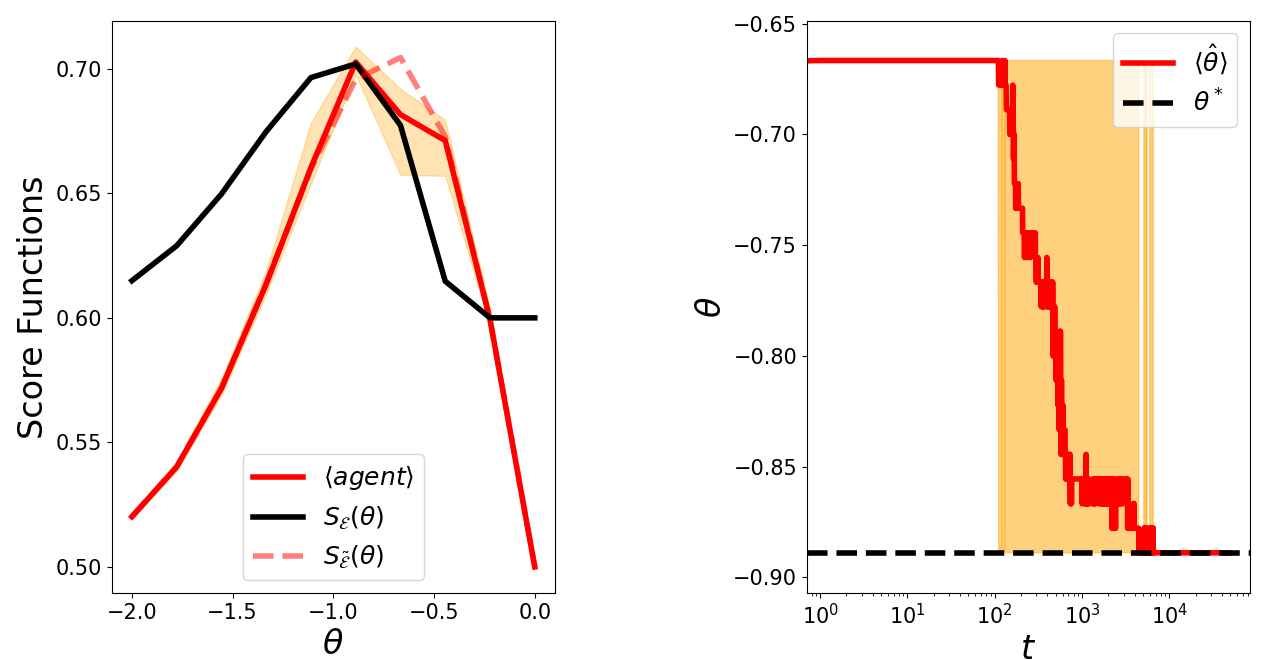}
    \caption{We show the average internal strategy of $25$ calibrating agents to fine-tune the device configuration under a change-of-prior scenario. Specifically, we depict the $Q$-values (left panel) from the effective model, and the ones obtained after RL fine-tuning. In the right panel we show the evolution of agent’s greedy strategy.}
\label{fig:change_priors_arecali}
\end{figure}

\textit{Q-learning Hyperparameters}. We scheduled the exploration rate of $\pi_\epsilon$ as per $\epsilon_E = \max\left(\epsilon_0, \epsilon_E \Delta_\epsilon\right)$, with $\epsilon_0, \Delta_\epsilon\in[0, 1)$ and $\epsilon_{t=0}=1$, \textit{e.g.} it is reduced over different episodes. Here, $\epsilon_0$ provides the minimum exploration level, while $\Delta_\epsilon$ gives a rate of change in the exploration. On a different note, we slightly modify the uniform sampling in $\pi_\epsilon$ by $p(\hat{a}|\hat{Q}) = \frac{1}{\mathcal{N}}\exp\left(-\Delta\left|\hat{Q}\left(\hat{a}^{*}\right) - Q(\hat{a})\right|^2\right)$; where $\hat{Q}(a)$ is the current Q-value estimate, $\hat{a}^*$ its associated greedy action, and $\Delta$ an importance-sampling parameter ($\Delta=0$ returns a uniformly random distribution). Finally, the Q-learning learning-rate $\lambda_E$ in Eq.~\ref{eq:QLUPDATERULE} is set decay as $\sim 1/E$. Note that  because of the initial information obtained by setting the effective model, we allow the learning-rate to take smaller values as per $\frac{1}{t+ \Delta_l}$, where $\Delta_l$ can be understood as how much the effective model is trusted by the RL agent.

All the hyperparameters used in the implementation can be observed in Table~\ref{table:hyperparams}.

\begin{figure}[b!]
\centering
\includegraphics[scale=0.27]{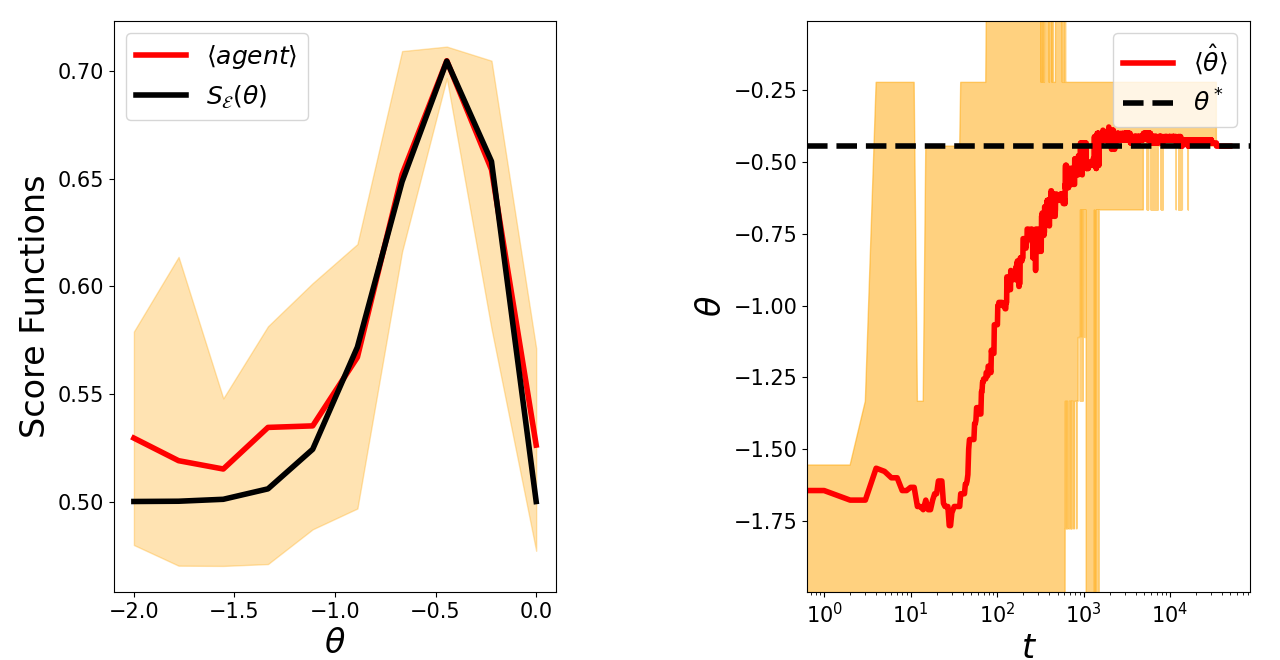}\\
    \includegraphics[scale=0.27]{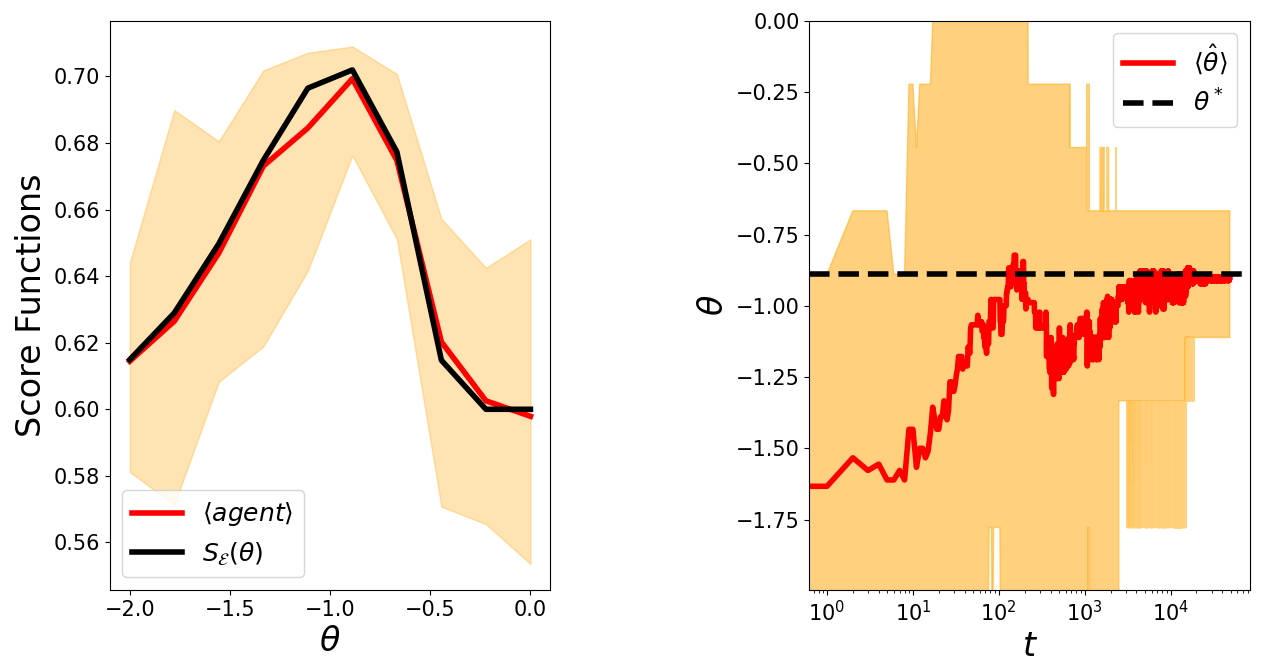}
    \caption{\textit{Re-calibration learning curve}. We show the mean internal strategy of $25$ reinforcement-learning agent’s, to optimize the device configuration, \textit{(top)} using a faulty displacement and \textit{(bottom)} a change of the prior probability. Specifically, we depict the Q-values (left panel) obtained by trial and error and the evolution of the greedy strategy (right panel).}
    \label{fig:change_betas_Q}
\end{figure}

\textit{Change of priors example}. To test the resilience of the method to a different noise source, we now study the situation in which a change in the prior probability $p_k$ of sending the classical bit $k$ gets modified as per $p_k \rightarrow p_k(\lambda_2) = \frac{1}{2} + (-1)^k \lambda_2$,
where $\lambda_2$ stands for the noise parameter, unknown by the agent. The results of our automatic re-calibration method for this scenario are show in in Fig. \ref{fig:change_priors_arecali}. Here, we average the results over $25$ instances obtained through random initializations, taking $~10^3$ states to find the optimal configuration in $\%90$ of the runs and $7 10^3$ to finish.

\textit{Comparison with Q-learning:} Finally, let us compare our techniques with traditional Q-learning method~\cite{bilkis2020real}. The results are benchmarked in Fig~\ref{fig:change_betas_Q}, under noisy scenarios previously considered. As observed, traditional Q-learning presents higher fluctuations over the Q-value estimates, requiring $~10x$ the amount of experiments than the method introduced in this paper, which exploits the usage of an effective model. 

\end{document}